\documentclass[11pt,tightenlines,aps,prd,groupedaddress,nofootinbib,longbibliography,notitlepage]{revtex4-1}

\usepackage{amsmath,amssymb,bm,mathtools}

\usepackage{graphicx}
\usepackage{amsfonts}
\usepackage{amssymb}
\usepackage{amsbsy}
\usepackage{amsmath}
\usepackage{latexsym}
\usepackage{bm}
\usepackage{color}
\usepackage{hyperref}
\usepackage{comment}
\usepackage[normalem]{ulem}
\usepackage[makeroom]{cancel}

\usepackage{stackengine}
\stackMath
\newcommand\tenq[2][1]{%
 \def\useanchorwidth{T}%
  \ifnum#1>1%
    \stackon[0pt]{\tenq[\numexpr#1-1\relax]{#2}}{\scriptscriptstyle\sim}%
  \else%
    \stackon[1pt]{#2}{\scriptscriptstyle\sim}%
  \fi%
}

\newcommand{\eq}[1]{(\ref{#1})}

\def\p {\partial}

\def\nn {\nonumber}

 \def\betad{\dot{\beta}}
 \def\phid{\dot{\phi}}
\def\eps{\epsilon}

\newcommand{\de}{\mbox{d}}
\newcommand{\be}{\begin{equation}}
\newcommand{\ee}{\end{equation}}
\newcommand{\pha}{\phantom{a}}

\newcommand{\pa}{\partial}
\newcommand{\Lie}{\mathcal{L}}
\newcommand{\bea}{\begin{eqnarray}}
\newcommand{\eea}{\end{eqnarray}}

\newcommand{\mar}[1]{{\color{blue}{{\bf MdC}: [ #1]}}}

\numberwithin{equation}{section}

\renewcommand{\theequation}{\arabic{section}.\arabic{equation}}

\begin{document}

\title{Physical Hamiltonian for mimetic gravity}

\author{Marco de Cesare}
\email{marco.de\_cesare@unb.ca}

\author{Viqar Husain}
\email{vhusain@unb.ca}

\affiliation{Department of Mathematics and Statistics\\
University of New Brunswick\\
Fredericton, NB E3B 5A3, Canada}

%\date{\today}

\begin{abstract}
 
Starting from a local action for mimetic gravity that includes higher derivatives of a scalar field $\phi$,
we derive a gauge-fixed canonical action of the theory in the ADM canonical formalism in the time gauge  $\phi=t$. This reduced action reveals
(i) a non-vanishing conserved physical Hamiltonian  that is a sum of two terms, the expression for the Hamiltonian constraint of general relativity and a function of the expansion scalar, and (ii) a reduced symplectic structure that geometrically  provides the Dirac brackets.  As applications of our general analysis, we compute the physical Hamiltonians and canonical equations for perturbations around Minkowski spacetime, homogeneous cosmologies, and spherically symmetric spacetimes.
\end{abstract}

\maketitle

\nopagebreak

\section{Introduction}

Since there is no widely accepted final theory of quantum gravity from which phenomenological consequences may be computed,  there is continuing interest in effective classical theories that are postulated to include expected effects from quantum gravity.  One such feature is singularity avoidance in cosmology and black-hole physics. Among such theories  are effective loop quantum cosmology~\cite{Ashtekar:2011ni}, Ho{\v r}ava-Lifshitz gravity \cite{Horava:2009uw},  and most recently,  mimetic gravity \cite{Chamseddine:2013kea,Chamseddine:2016uef}.

While the main interest in such models is often restricted  to derivation of special classes of solutions such as cosmological models and spherically symmetric geometries, it is also useful to study their general structure, particularly from the canonical point of view. This is because the manifestation of general covariance in Hamiltonian theories is through the algebra of first class constraints. For pure gravity theories with only metric degrees of freedom, there is a severe restriction on possible deformations of constraints \cite{Hojman:1976vp}:  first class constraints obey the Dirac-Bergmann algebra of hypersurface deformations. Inclusion of matter fields allows more possibilities, permitting certain modified algebras. Examples of this are the use of matter reference systems  Ref.~\cite{Brown:1994py} and anomaly-free deformations of the constraints algebra designed to encode quantum-geometry corrections~\cite{Bojowald:2011aa}.

 For gravity theories the canonical formulation is the appropriate framework to identify convenient choices of time and their corresponding physical Hamiltonians. This is potentially useful not just for quantization and the related `problem of time', but also for understanding features of the dynamics at both the classical and  quantum levels ---some physical Hamiltonians may be   more useful than others, particularly  if they turn out to be time-independent. This last feature typically requires matter time gauges rather than geometric ones made from the Arnowitt-Deser-Misner (ADM) variables \cite{Brown:1994py}; obvious choices such as $t=\mbox{(spatial volume)}$ yield unwieldly time-dependent Hamiltonians that are singular at $t=0$.  For GR coupled to pressureless irrotational dust \cite{Brown:1994py,Husain:2011tk,Giesel:2012rb}, or a massless scalar field with zero potential \cite{Rovelli:1993bm,Alesci:2015wla}, the $\phi=t$ gauge condition yields Hamiltonians that are time-independent.  As we show, this gauge is also a natural one in scalar-tensor theories such as mimetic gravity, although in the literature on these models $\phi=t$ is usually seen as a consequence of the field equations in synchronous coordinates (see, e.g., Ref.~\cite{Chamseddine:2016uef}), rather than as a canonical gauge in the Hamiltonian theory.  It is with this perspective in mind that we approach the topic of this paper.
 
Our main result in this paper is a derivation of the physical Hamiltonian for mimetic gravity in the time gauge $\phi=t$ ( `dust time gauge').  This follows a path similar to the derivation for GR with dust \cite{Husain:2011tk}, but has a certain distinctive feature; this is a restriction of the symplectic structure to a surface in the phase space that goes beyond just the condition  $\phi=t$ due to presence of auxiliary fields in the starting action. We note that a Hamiltonian analysis of mimetic gravity appeared in Refs.~\cite{Kluson:2017iem} and \cite{Bodendorfer:2017bjt}.  However neither of these works  considers $\phi=t$ as a canonical gauge choice that naturally provides a reduced action, symplectic structure and physical Hamiltonian. 
 
In the present work we focus on the version of mimetic gravity proposed in Ref.~\cite{Chamseddine:2016uef}, using the simpler equivalent action used in 
Ref.~\cite{Bodendorfer:2017bjt}.  This theory is a generalization of the original mimetic gravity~\cite{Chamseddine:2013kea} (whose reformulation \cite{Golovnev:2013jxa} led to further developments Refs.~\cite{Chamseddine:2014vna,Chamseddine:2016uef}).
For a review, see Ref.~\cite{Sebastiani:2016ras}.

Theories of this type and their generalizations have been applied in various contexts, including cosmological models, where the question of whether its equations correspond to those of effective loop quantum cosmology (LQC) is addressed \cite{Bodendorfer:2017bjt, Liu:2017puc,deCesare:2019pqj,deCesare:2018cts,deHaro:2018sqw}. It has also been noted that the mimetic gravity models belong to a  class of modified gravity theories where the Kasner exponents in the pre- and post- bounce phases obey the same transition rules as in LQC \cite{deCesare:2019suk,Wilson-Ewing:2017vju}. However these applications to LQC have limitations in the anisotropic \cite{Bodendorfer:2018ptp,deCesare:2020swb} and spherically symmetric sectors \cite{BenAchour:2017ivq}.

The outline of the paper is as follows. In Section~\ref{Sec:Canonical} we give the canonical analysis  of the action  in the ADM formalism. This begins with the action given in Ref.~\cite{Bodendorfer:2017bjt}, but differs in the subsequent analysis. We show that the Hamiltonian and diffeomorphism constraints are first class and close as the standard Dirac-Bergmann algebra. We use this fact in Sec.\ref{Sec:DustTimeGauge} to fix the gauge $\phi=t$, show that it is free from the Gribov ambiguity, and proceed to derive the reduced canonical action. This requires a reduction of the symplectic structure to take into account the field equations of the auxiliary fields in the action.
 In Sec.~\ref{Sec:Linearized} we analyze the linearized theory around Minkowski spacetime. In Sec.~\ref{Sec:SymmetryReduced} we apply the general results to the flat FLRW model and to spherically symmetric spacetimes. We conclude in Sec.~\ref{Sec:Discussion} with a summary and discussion. Finally in a technical appendix (Appendix~\ref{Sec:Appendix})  we analyze in detail a singular limit of the theory and show that this limit has hidden symmetries. (We use the  metric signature $(-+++)$, and units such that $c=8\pi G=1$.)

\section{Canonical analysis}\label{Sec:Canonical}
   We begin with the action \cite{Bodendorfer:2017bjt}
    \begin{align}\label{action}
S[g_{ab},\phi,\lambda,\beta,\chi]&=\int_{\Sigma\times R} \de^4 x\sqrt{-g}\left[\frac{R}{2}-\frac{\lambda}{2}\left(1+g^{ab}\pa_a\phi\pa_b\phi\right)+f(\chi)+\beta\chi-g^{ab}\pa_a\beta\pa_b\phi\right] \nn\\ \nn\\
&\equiv S^G [g_{ab}]+ S^M[\phi,\lambda,\beta,\chi,g_{ab}],
   \end{align}
   where $S^G$ denotes the Einstein-Hilbert action and $S^M$ is the action for the scalar field sector.\footnote{Our conventions for the signs of the  terms in the action \eqref{action} are slightly different from Ref.~\cite{Bodendorfer:2017bjt}, and are such that on-shell we have $\chi=-\Box\phi$, that is consistent with the geometric interpretation of $\chi$ as the expansion in the synchronous gauge \cite{Chamseddine:2016uef}.} 
   The spacetime manifold is assumed to be $\Sigma\times\mathbb{R}$. 
   The action~\eqref{action} is dynamically equivalent to the version of mimetic gravity proposed in Ref.~\cite{Chamseddine:2016uef}  (which includes higher derivatives of the scalar $\phi$ through the function $f$\;), but is better suited for a canonical analysis. (The original action of Ref.~\cite{Chamseddine:2016uef} is recovered by eliminating the auxiliary fields $\beta$ and $\chi$, using their equations of motion.)
   
 We start by reviewing the ADM decomposition of the action~\eqref{action} \cite{Bodendorfer:2017bjt}, generalizing the well-known procedure in GR.
We introduce on the manifold a time-like vector field $t^a = Nn^a + N^a$\,, where $n^a$ is the unit normal to the spatial hypersurfaces $\Sigma$\,. 
 This leads to the  definition of the positive-definite spatial metric $q_{ab} = g_{ab} +n_an_b$\,, and 
 \be
 \sqrt{-g} = N\sqrt{q}~, \quad g^{ab} = q^{ab} - \frac{1}{N^2} ( t^a-N^a)(t^b-N^b)~.
 \ee  
%The ADM Hamiltonian decomposition of the GR action (i.e.,~the first term in \eqref{action}) is well-known~\cite{Arnowitt:1962hi}.
%
%
%
%
%
The action can then be rewritten, up to a boundary term, as
\be
S=\int \de^4 x\; N\sqrt{q}\left[\frac{R}{2}-\frac{\lambda}{2}\left(1+q^{ab}\p_a\phi \p_b\phi -(\Lie_n\phi)^2\right)+f(\chi)+\beta\chi-q^{ab}\p_a\beta \p_b\phi+\Lie_n\beta\Lie_n\phi\right]~.
\ee
The Lie derivative of a generic scalar function $\mathcal{F}$ along the normal direction $n^a$ can be decomposed as follows
\be\label{Lie_decomposition}
\Lie_n \mathcal{F}=n^a\pa_a \mathcal{F}=\frac{1}{N}\left(\dot{\mathcal{F}}-\Lie_N \mathcal{F}\right) ~,
%\Lie_n\beta=n^a\pa_a\beta=\frac{1}{N}\left(\dot{\beta}-N^a\pa_a\beta\right)~,\\
%\Lie_n\phi=n^a\pa_a\phi=\frac{1}{N}\left(\dot{\phi}-N^a\pa_a\phi\right)~.
\ee
where an overdot is used to denote the Lie derivative along $t^a$, i.e.~$\dot{\mathcal{F}}\coloneqq \Lie_t \mathcal{F}$\,.
Using Eq.~\eqref{Lie_decomposition}, we can easily obtain the canonical momenta expressed in terms of the velocities
\begin{align}\label{momenta_beta_phi}
p_\beta&\coloneqq\frac{\delta S}{\delta\dot{\beta}}=\sqrt{q}\,\Lie_n\phi=\frac{\sqrt{q}}{N}\left(\dot{\phi}-\Lie_N\phi\right)~,\\
p_\phi&\coloneqq\frac{\delta S}{\delta\dot{\phi}}=\sqrt{q}\,\left(\lambda \Lie_n\phi+\Lie_n\beta\right)= \frac{\sqrt{q}}{N}\left[(\lambda\,\dot{\phi}+\dot{\beta})-(\lambda\,\Lie_N \phi+\Lie_N\beta)\right]~.
\end{align}
The momenta canonically conjugated to $\chi$, $\lambda$, $N$, and $N^a$ vanish identically. Since the gravitational sector of \eqref{action} is the same as in GR, the relation between the canonical momentum $\pi^{ab}$ and the extrinsic curvature $K_{ab}$ is the standard one
\be\label{momentum_geometry}
\pi^{ab}\coloneqq\frac{\delta S}{\delta\dot{q}_{ab}}=\frac{\sqrt{q}}{2}(K^{ab}-K q^{ab})~,
\ee 
where the extrinsic curvature is defined as usual
\be
K_{ab}\coloneqq\frac{1}{2}{\cal L}_n q_{ab}=\frac{1}{2N}\left(\dot{q}_{ab}-2D_{(a}N_{b)}\right)~.
\ee

Inverting the relations \eqref{momenta_beta_phi} and \eqref{momentum_geometry} and substituting for the velocities in \eqref{action} gives the canonical action
\be\label{canonical_action1}
S=\int \de t\,\de^3x \left( \pi^{ab}\dot{q}_{ab}+p_\phi\dot{\phi}+p_\beta\dot{\beta}-N\mathcal{H}-N^a \mathcal{C}_a \right)~.
\ee
where %the Hamiltonian and diffeomorphism constraints are respectively given by
\be\label{Hamiltonian constraint}
\begin{split}
\mathcal{H}=&\frac{2}{\sqrt{q}}\left(\pi_{ab}^2-\frac{1}{2}\pi^2\right)-\frac{\sqrt{q}}{2}R^{(3)}+\frac{p_\beta}{\sqrt{q}}\left(p_\phi-\frac{\lambda}{2}p_\beta\right)\\
&+\sqrt{q}\left[\frac{\lambda}{2}\left(1+q^{ab}D_a\phi D_b\phi\right)+q^{ab}D_a\beta D_b \phi-f(\chi)-\beta\chi\right]~,
\end{split}
\ee
\be\label{eq:diffeo_constraint}
\mathcal{C}_a=-2D_b\pi^b_{\pha a}+p_\phi D_a\phi+p_\beta D_a\beta ~.
\ee
%Defining the canonical momenta $p_\beta,\p_\phi$ and substituting for the velocities gives the canonical matter action  
%\be
%S^M = \int dtd^3x \left[\betad p_\beta + \phid p_\phi -H^M - C^M \right]
%\ee
%where 
%\begin{align}
%H^M &\equiv \frac{N}{\sqrt{q}}\ p_\beta p_\phi + N\sqrt{q} \left[  q^{ab} \p_a\beta\p_b\phi +    \frac{\lambda}{2} \left(1+ q^{ab} \p_a\phi\p_b\phi  -\frac{p_\beta^2}{q}  \right)  +\beta\chi + f(\chi)   \right] \nn\\
%C^M &\equiv p_\beta{\cal L}_N\beta + p_\phi{\cal L}_N\phi.
%\end{align}
%Therefore the full canonical theory including gravity is 
%\be
%S = \int dt d^3x \left[ \pi^{ab} \dot{q}_{ab} + \betad p_\beta + \phid p_\phi  - H^G -H^M -C^G-C^M \right]
%\ee
 %where $H^G$ and $C^M$ are the terms in the Hamiltonian for general relativity.
 %
% Variation of this action with respect to $N$ and $N^a$ gives the constraints.  These constraints remain first class. 
 %
 %
 Varying the action w.r.t. $N$ and $N^a$ gives the Hamiltonian and diffeomorphism constraints
\be
\mathcal{H}\approx 0~,\qquad \mathcal{C}_a\approx 0~.
\ee

At this stage  we do not derive the remaining   constraints by applying the Dirac algorithm (as done in Ref.~\cite{Bodendorfer:2017bjt}). Instead, as we will see below,  it is technically  advantageous to first fix a canonical time gauge after establishing that the surface deformation algebra remains first class.
      
%%%%%%%%%
\iffalse
Variation w.r.t.~the Lagrange multiplier $\lambda$ and the auxiliary field $\chi$ yields the following constraints
\begin{align}
C_{\lambda}&\coloneqq p_{\beta}^2-q\left(1+q^{ab}D_a\phi D_b\phi\right)\approx 0~,\label{PbetaConstraint}\\
C_{\chi}&\coloneqq \beta+f^{\prime}(\chi)\approx0~.\label{betaConstraint}
\end{align}
The constraint (\ref{PbetaConstraint}) can be solved for $p_\beta$
\be
p_{\beta}\approx \pm \sqrt{q}\sqrt{1+q^{ab}D_a\phi D_b\phi} ~.
\ee
\fi
%%%%%%%%%

%The canonical action \eqref{canonical_action1} contains constns information on the Hamiltonian dynamics of the theory.
%In Section~\ref{Sec:ConstraintsAlgebra} we will show that the Hamiltonian and diffeomorphism constraints are first class, as in GR. In %particular, since the Hamiltonian constraint is first class, the physical degrees of freedom of the theory can be identified by means of a %convenient gauge-fixing of the time-reparametrisation gauge freedom, as we will show in Section~\ref{Sec:DustTimeGauge}.

%We remark that, for the purpose of analysing the dynamics of the gauge-fixed theory, it is not necessary to derive the second class constraints %at this stage.\footnote{Second class constraints for the theory at hand have been obtained in Ref.~\cite{Bodendorfer:2017bjt} following the %standard Dirac algorithm.} In fact, the canonical action {\it per se} is enough to ensure the consistency of the dynamics. It is also technically %much easier to derive and solve the second class constraints starting from the gauge-fixed action, as we shown in %Section~\ref{Sec:DustTimeGauge}.

\subsection{Contraint algebra}\label{Sec:ConstraintsAlgebra}

We now show that the algebra of constraints algebra is the expected Dirac-Bergmann algebra, and is therefore first class. This necessary step sets the stage for fixing the time gauge $\phi=t$, which we carry out in the next section. 

The non-trivial calculation is the Poisson bracket of the Hamiltonian constraint with itself.
It is convenient to split the Hamiltonian constraint into the sum of two terms $\mathcal{H}=\mathcal{H}^{\rm G}+\mathcal{H}^{\rm M}$, representing a gravitational contribution and non-standard matter Hamiltonian, given respectively by 
\begin{align}
\mathcal{H}^{\rm G}=&\frac{2}{\sqrt{q}}\left(\pi_{ab}^2-\frac{1}{2}\pi^2\right)-\frac{\sqrt{q}}{2}R^{(3)}~,\\
\mathcal{H}^{\rm M}=&\frac{p_\beta}{\sqrt{q}}\left(p_\phi-\frac{\lambda}{2}p_\beta\right)+\sqrt{q}\left[\frac{\lambda}{2}\left(1+q^{ab}D_a\phi D_b\phi\right)+q^{ab}D_a\beta D_b \phi-f(\chi)-\beta\chi\right]~.
\end{align}
 $\mathcal{H}^{\rm G}$ coincides with the standard GR Hamiltonian constraint, therefore we have the standard result 
 \be
\{\mathcal{H}^{\rm G}(x),\mathcal{H}^{\rm G}(y)\}=\big(2\pi_{ab}(x)-\pi(x) q_{ab}(x)\big) \hat{{\cal T}}_{(y)}^{ab}\delta(x,y) -\left(x\leftrightarrow y\right)~,
\ee
where  
\be
\hat{{\cal T}}^{ab}\coloneqq 2D^{(a}D^{b)}-q^{ab}D^cD_c~.
\ee
 The Poisson bracket for the matter terms is   
\be
\{\mathcal{H}^{\rm M}(x),\mathcal{H}^{\rm M}(y)\}=-q^{ab}(y)\big(p_{\phi}(x)D_a\phi(y)+p_{\beta}(x)D_a\beta(y)\big)D_b^{(y)}\delta(x,y) - \left(x\leftrightarrow y\right),
\ee
and for the the mixed term it is 
\be\label{Eq:PB_gravity_matter}
\{\mathcal{H}^{\rm G}(x),\mathcal{H}^{\rm M}(y)\}\propto \delta(x,y) ~.
\ee
The detailed form of the proportionality factor in Eq.~\eqref{Eq:PB_gravity_matter} is unimportant for our purposes; it will be sufficient to note that no derivatives of the delta-function appear in \eqref{Eq:PB_gravity_matter}, which implies that such a term is exactly cancelled by $\{\mathcal{H}^{\rm M}(y),\mathcal{H}^{\rm G}(x)\}$. Combining these results gives  the Poisson bracket of the full Hamiltonian constraint with itself, 
\be
\begin{split}
&\{\mathcal{H}(x),\mathcal{H}(y)\}=\{\mathcal{H}^{\rm G}(x),\mathcal{H}^{\rm G}(y)\}+\{\mathcal{H}^{\rm M}(x),\mathcal{H}^{\rm M}(y)\}=\\
&\big(2\pi_{ab}(x)-\pi(x) q_{ab}(x)\big) \hat{T}_{(y)}^{ab}\delta(x,y)-q^{ab}(y)\big(p_{\phi}(x)D_a\phi(y)+p_{\beta}(x)D_a\beta(y)\big)D_b^{(y)}\delta(x,y)-\left(x\leftrightarrow y\right) ~.
\end{split}
\ee

Denoting the smeared Hamitonian and diffeomorphism constraints respectively as $H^{\perp}[N]=\int \de^3 x\; N(x)\mathcal{H}(x)$ and $C[\vec{N}]=\int \de^3 x\; N^a\mathcal{C}_a(x)$, we obtain from the above results
\be
\{H^{\perp}[M],H^{\perp}[N]\}=C[\vec{V}]~,
\ee
where $V^a=h^{ab}(M\pa_b N-N\pa_b M)$.
The remaining Poisson brackets are straightforward to compute, since the vector constraint \eqref{eq:diffeo_constraint} is canonical and therefore is a generator of the algebra of three-dimensional diffeomorphisms. A straightforward standard calculation gives
\begin{align}
\{C[\vec{M}],C[\vec{N}]\}=C[\mathcal{L}_{\vec{N}}\vec{M}]\\
\{C[\vec{M}],H^{\perp}[N]\}=H^{\perp}[\mathcal{L}_{\vec{N}}N].
\end{align}

%This proves that the Hamiltonian and diffeomorphism constraints in this theory close the standard Dirac-Bergmann algebra, with the same %structure functions as in GR, and with no anomalies. This is consistent with the findings in Ref.~\cite{Bodendorfer:2017bjt}. 

 \section{`Dust time gauge' and reduced canonical action}\label{Sec:DustTimeGauge}
 
 Having established that  the Hamiltonian constraint is first class, that is, the time-reparametrizations it generates are gauge transformations, we can proceed to identify the physical degrees of freedom  by  a canonical gauge-fixing of this transformation. This amounts to setting a scalar function of phase space variables to be time; the negative of the canonically conjugate phase space function is then the physical Hamiltonian.  After the gauge fixing, we use the equations of motion for the canonical variables $\beta$ and $p_\beta$ to reduce the theory further to obtain a final action of only the ADM variables $(q_{ab},\pi^{ab})$. These steps form the key differences from the procedure followed in  \cite{Bodendorfer:2017bjt}.
 
 \subsection{Time gauge fixing}
We impose the canonical gauge condition $\phi=t$ for the time coordinate. This is a good gauge-fixing since it is second class with the Hamiltonian constraint;  denoting $\mathcal{G}\coloneqq \phi-t$ we have 
\be\label{gauge_condition_PB}
\{\mathcal{G}(x),\mathcal{H}(y)\}=\frac{p_\beta}{\sqrt{q}}\,\delta(x,y)~,
\ee
 and the Dirac matrix 
\be
\Delta=\left(\begin{array}{cc} 0&\{\mathcal{G},\mathcal{H}\}\\ \{\mathcal{H},\mathcal{G}\} &0\end{array}\right)=\frac{p_\beta}{\sqrt{q}}\left(\begin{array}{cc} 0&1\\ -1 &0\end{array}\right)
\ee
is everywhere non-degenerate, with the exception of the points where $p_\beta/\sqrt{q}\neq0$. But these points are not realized dynamically since $p_\beta\ne0$ and $\sqrt{q} \rightarrow \infty$ is not realized in finite time.

 Locally, Eq.~\eqref{gauge_condition_PB} means that the gauge orbits intersect the gauge-fixing surface $\mathcal{G}=0$ once and only once. This is true also globally (i.e.~there is no Gribov ambiguity) since the Faddeev-Popov determinant is non-zero everywhere; this determinant   is given by the Pfaffian of the Dirac matrix $\Delta$ (see e.g. Ref. \cite{Faddeev_1969}), and therefore equals $p_\beta/\sqrt{q}$.
 
To obtain the gauge fixed action we solve the Hamiltonian constraint ${\cal H}^{\rm G} + {\cal H}^{\rm M}=0$ strongly for $p_\phi$ to get 
 \be\label{Pphi_solution}
 p_\phi=-\frac{\sqrt{q}}{p_\beta}\,\mathcal{H}^{\rm G}+\frac{q}{p_\beta}\left[f(\chi)+\beta\chi-\frac{\lambda}{2}\left(1-\frac{p_\beta^2}{q}\right)\right]~.
\ee
 We also have the condition that the gauge be preserved in time
 \be
 1=\phid = \left\{ \phi, \int d^3x \, (N\mathcal{H}^{\rm M}+N^a \mathcal{C}^{\rm M}_a) \right\}_{\phi=t}   = \left[\frac{Np_\beta}{\sqrt{q}} + {\cal L}_N\phi \right]_{\phi=t}.
 \ee
 This fixes the lapse function
 \be\label{lapse_solution}
 N = \frac{\sqrt{q}}{p_\beta}~.
 \ee 
 Substituting Eqs.~\eqref{Pphi_solution} and \eqref{lapse_solution}, and the gauge condition $\phi=t$ into the canonical action, we obtain the gauge fixed action 
 \be\label{GFaction29}
S^{\rm GF}[q,\pi,\beta,p_\beta,\chi,\lambda]  = \int \de t\, \de^3x \left[ \pi^{ab} \dot{q}_{ab} +  p_\beta  \betad - \tilde{\cal H} - N^a({\cal C}_a^{\rm G} + {\cal C}_a^\beta)\right]~,
\ee
 where
 \be\label{Hphysical0}
 \tilde{\cal H}= \frac{\sqrt{q}}{p_\beta} {\cal H}^{\rm G} - \frac{q}{p_\beta}\left[ f(\chi)+ \beta\chi -\frac{\lambda}{2} \left(1- \frac{p_\beta^2}{q} \right)\right] ~.
 \ee
 This expression is a function of the canonical pairs $(q_{ab},\pi^{ab})$ and $(\beta,p_\beta)$, and the auxiliary fields $\lambda$ and $\chi$. We note that Eq.~\eqref{Hphysical0} represents a true Hamiltonian density, as opposed to the Hamiltonian constraint, and the diffeomorphism constraint remains as a the only gauge symmetry.

\subsection{Elimination of auxiliary fields} 
 
 At this stage we would like to eliminate the auxiliary fields $\beta$, $p_\beta$, and $\chi$. To begin with, we note that variation of the action~\eqref{GFaction29} w.r.t.~$\lambda$ and $\chi$ respectively leads to the following   equations
 \be\label{Constraints_AuxiliaryFields}
 {\cal C}_{\lambda}\coloneqq p_\beta^2 -q=0~,\quad  {\cal C}_{\chi}\coloneqq\beta+f^{\prime}(\chi)=0~.
 \ee
 The constraint ${\cal C}_{\lambda}=0$ shows that $N=  \sqrt{q}/p_\beta=1$.
 Variation of~\eqref{GFaction29} w.r.t.~$\beta$ gives, using~\eqref{Constraints_AuxiliaryFields}
\be\label{chi_expansion}
\chi = \frac{1}{\sqrt{q}} \left(  {\cal L}_t-{\cal L}_{\vec N}\right) \sqrt{q}={\cal L}_n \ln\sqrt{q}~.
\ee
Using the equation of motion 
\be
{\cal L}_t \sqrt{q} = \left\{ \sqrt{q} ,  \int \de^3x \left[ {\cal H}^{\rm G} + N^a {\cal C}^{\rm G}\right] \right\}=-\pi+{\cal L}_N \sqrt{q}~,
\ee
(with $\pi\coloneqq \pi^{ab}q_{ab}$) then gives 
\be\label{chi_momentumtrace}
\chi = %- \frac{1}{\sqrt{q}} \left\{ \sqrt{q} ,  \int d^3x\ {\cal H}^G\right\}= -\pi^{ab}q_{ab} \equiv
-\frac{\pi}{\sqrt{q}}~.
\ee
Equation \eqref{chi_expansion} shows that $\chi$ admits a neat geometric interpretation as the expansion of the congruence generated by the normal vector field $n^a$ in the $\phi=t$ gauge. We also note that $\chi$ is proportional to the momentum conjugate to the volume $V=\sqrt{q}$.

 The constraints~\eqref{Constraints_AuxiliaryFields} determine a surface in the time-gauge fixed phase space with canonical coordinates
 $(q_{ab},\pi^{ab},\beta,p_\beta)$. These  constraints  have three effects on the action: they (i) simplify   the physical Hamiltonian~\eqref{Hphysical0}  to a function of only the ADM variables
\be\label{physical_Hamiltonian_final}
{\cal H}^{\rm P}= {\cal H}^{\rm G} + \sqrt{q}\left( \chi\, f^{\prime}(\chi) -f(\chi)\right) ~,\quad \mbox{with}~~\chi =-\frac{\pi}{\sqrt{q}}~,
\ee
and (ii)  modify the symplectic potential to  \cite{Barnich:1991tc}
\bea
\omega&=&\int \de^3 x\; \left( \pi^{ab} \delta q_{ab}+p_\beta\delta\beta\right)_{{\cal C}_{\lambda},{\cal C}_{\chi}=0}=\int \de^3 x\;\left(\pi^{ab} \delta q_{ab}-\beta\delta p_\beta\right)_{{\cal C}_{\lambda},{\cal C}_{\chi}=0} \nn\\
&=&\int \de^3 x\;\left(\pi^{ab}+f^{\prime}(\chi)\frac{\sqrt{q}}{2}q^{ab} \right)\delta q_{ab}, 
\label{symplectic_potential_reduced}
\eea
where $\delta$ is the exterior derivative on phase space and the  second equality holds up to an exact one-form, and (iii) modify the diffeomorphism constraint to 
\be
\bar{C}_a\equiv \left(-2D_b\pi^b_{\ a} + p_\beta D_a\beta\right) _{{\cal C}_{\lambda},{\cal C}_{\chi}=0}
= -2 D_b\left(\pi^b_{\ a} + \frac{f'(\chi)}{2}\sqrt{q}q^b_{\ a} \right)=0.
\label{newdiff}
\ee
The last two equations (\ref{symplectic_potential_reduced}-\ref{newdiff}) show respectively that the new momentum 
\be
\bar{\pi}^{ab} := \pi^{ab}  + \frac{f'(\chi)}{2}\sqrt{q}q^{ab}
\label{new_momentum}
\ee
is canonically conjugate to $q_{ab}$ on the surface defined by (\ref{Constraints_AuxiliaryFields}), and the diffeomorphism constraint becomes
\be
\bar{{\cal C}}_a\equiv -2D_b\bar{\pi}^b_{\ a}=0.
\ee
This form makes it clear  that this constraint remains first class. Putting these results together the action simplifies to 
\be\label{GFaction_final}
S^{\rm GF}[q,\bar{\pi}]  = \int \de t\, \de^3x \left[ \bar{\pi}^{ab}  \dot{q}_{ab} - \bar{{\cal H}}^{\rm P} - N^a \bar{{\cal C}}_a \right]~,
\ee
where 
\be
\bar{{\cal H}}^{\rm P} = \frac{2}{\sqrt{q}}\left(\bar{\pi}_{ab} \bar{\pi}^{ab}-\frac{1}{2}\bar{\pi}^2\right)-\frac{\sqrt{q}}{2}R^{(3)}~ - \sqrt{q}\left( f(\chi)-\frac{3}{4} \left(f^{\prime}(\chi)\right)^2 \right), 
\label{HPbar}
\ee
and  $\chi$ and $\bar{\pi}$ are related using (\ref{chi_momentumtrace}) and (\ref{new_momentum}) by 
\be
\bar{\pi}=\sqrt{q}\left(\frac{3}{2}f^{\prime}(\chi)-\chi\right).
\label{pibar-chi}
\ee
 This action and Hamiltonian constitute our main result. We note that the action may also be written in terms of the original ADM variables as
 
 \be\label{GFaction_final2}
S^{\rm GF}[q,\pi]  = \int \de t\, \de^3x \left[ \left(\pi^{ab} +f^{\prime}(\chi)\frac{\sqrt{q}}{2}q^{ab} \right) \dot{q}_{ab} - {\cal H}^{\rm P} 
- N^a  \bar{C}_a \right]~,
\ee
with $ {\cal H}^{\rm P}$  and $\bar{C}_a$ as in (\ref{physical_Hamiltonian_final}) and (\ref{newdiff}).

An accounting of physical degrees of freedom is immediate from  \eqref{GFaction_final}: the canonical pair $(q_{ab},\bar{\pi}^{ab})$ represents  a 12-dimensional phase space per space point, subject to the three first-class constraints $\bar{{\cal C}}_a=0$; therefore (with exception $f'(\chi) = 2\chi/3$ corresponding to $\bar{\pi}=0$, to be discussed below), there are three independent physical configuration degrees of freedom per point. Thus, compared to GR there is one extra local degree of freedom. Perturbatively, this corresponds to a propagating scalar mode, which has been studied in several works, see e.g.~Refs.~\cite{Firouzjahi:2017txv,Langlois:2018jdg}, as well as Refs.~\cite{Chamseddine:2014vna,Ramazanov:2016xhp,Ijjas:2016pad} for earlier works with $f(\chi)$ quadratic; we re-derive this below in the canonical theory. It is interesting to observe that the number of degrees of freedom is three also in the special case $f(\chi)=0$, which corresponds to GR minimally coupled to a dust fluid in the dust time gauge \cite{Husain:2011tk}; in this special case the perturbative dynamics of the scalar mode becomes ultra-local (i.e., there are no spatial gradients in the second-order action), see Refs.~\cite{Ali:2015ftw,Husain:2020uac}.

We note the following additional remarks concerning the above  procedure.
\begin{itemize}

\item The canonical symplectic two-form on the partially reduced phase-space obtained from \eqref{symplectic_potential_reduced} is
\be\label{symplectic_twoform}
\Omega=-\delta\omega=\int\de^3x\;\left(\delta^a_{(c}\delta^b_{d)}-\frac{1}{2}q^{ab}q_{cd}\,f^{\prime\prime}(\chi)\right)\delta q_{ab}\wedge \delta \pi^{cd}~.
\ee
The corresponding Poisson bracket  is obtained by inverting the tensor in brackets in \eqref{symplectic_twoform}. This  gives  the Dirac bracket 
\be\label{Dirac_bracket}
\{q_{ab},\pi^{cd}\}_{\star}= \delta^c_{(a}\delta^d_{b)}+\frac{f^{\prime\prime}(\chi)}{2-3 f^{\prime\prime}(\chi)}q_{ab}q^{cd}~,
\ee
 provided that $f^{\prime\prime}(\chi)\neq\frac{2}{3}$. (The singular case where this condition does not hold is discussed below.)
 The canonical equations of motion for the variables $q_{ab}$ and $\pi^{ab}$ are obtained by varying the action~\eqref{GFaction_final}; they read as
\be\label{HamiltonEQS_general}
\dot{q}_{ab}=\{q_{ab},{\cal H}^{\rm P}\}_{\star}~,\quad
\dot{\pi}^{ab}=\{\pi^{ab},{\cal H}^{\rm P}\}_{\star}~,
\ee
with ${\cal H}^{\rm P}$ as in (\ref{physical_Hamiltonian_final}). Using Eq.~\eqref{physical_Hamiltonian_final} and the fundamental Dirac bracket \eqref{Dirac_bracket} it is easy, if tedious, to show that the first equation in \eqref{HamiltonEQS_general} gives the standard relation between velocity and momentum, consistently with Eq.~\eqref{momentum_geometry}.
We also observe that the physical Hamiltonian ${\cal H}^{\rm P}$ is a first integral of the system, since it does not depend on $\phi$-time explicitly. The existence of such a first integral in the gauge-fixed theory stems from the shift-invariance of the original action~\eqref{action}.

\item It is clear from Eq.~\eqref{Dirac_bracket} that the two-form~\eqref{symplectic_twoform} is not invertible if $f^{\prime\prime}(\chi)=\frac{2}{3}$. With the form
\be
f(\chi)=c_0+c_1\chi+\frac{1}{3}\chi^2.
\label{specialf}
\ee
we find from (\ref{chi_momentumtrace}) and (\ref{new_momentum})  that 
\be
\bar{\pi}^{ab} = \pi^{ab}  - \frac{1}{3}\pi q^{ab} +\frac{c_1}{2} \sqrt{q}\, q^{ab}, \quad \bar{\pi} = \frac{3c_1}{2} \sqrt{q},
\label{pi_sing}
\ee
and the Hamiltonian (\ref{HPbar}) becomes
\be
{\cal H}^{\rm P}_{\rm sing} = \frac{2}{\sqrt{q}} \bar{\pi}_{ab} \bar{\pi}^{ab} -\frac{\sqrt{q}}{2}\left(R^{(3)}+2\, c_0\right);
\label{Hsing}
\ee
this is independent of $c_1$. It is evident that (\ref{pi_sing}) defines a surface in phase space so that  $\bar{\pi}$ and $\sqrt{q}$ are no longer  independent. This is manifested in the symplectic structure, which takes the form 
\bea
\label{singular_symplectic}
\Omega&=&-\delta\omega=\int \de^3 x\; \left(\delta^a_{(c}\delta^b_{d)}-\frac{1}{3}q^{ab}q_{cd}\right)\delta q_{ab}\wedge \delta \pi^{cd} \nn\\
&=&\int \de^3 x\; \delta \bar{q}_{ab}\wedge \delta \bar{\pi}^{ab}~,
\eea
 where  $\delta\bar{\pi}^{ab}$ is the variation of \eqref{pi_sing} and
\be
\label{Eq:TracelessPI_TildeQ}
 \delta \bar{q}_{ab}\coloneqq \delta q_{ab}-\frac{1}{3}q_{ab}\frac{\delta q}{q}~.  
 \ee
 This expression is obtained from contracting the term in brackets in \eqref{singular_symplectic} (which is a projector onto the subspace of traceless symmetric matrices) with $\delta q_{ab}$. The symplectic two-form~\eqref{singular_symplectic} can also be recast as
\be
\Omega=\int \de^3 x\; \left(\delta q_{ab}\wedge \delta \pi^{ab}-\frac{1}{3q}\delta q\wedge \delta\pi \right)~.
\ee
It is therefore evident that the dimension of the phase space is reduced by two (per point).
 
For $c_1=0$, we have $\bar{\pi}^{ab}=\pi^{\langle ab\rangle}$ (i.e. its traceless part). The case $c_1\neq0$ is related to $c_1=0$ by a canonical transformation, since the corresponding symplectic potentials differ by an exact one-form:
\be
\bar{\pi}^{ab} \delta q_{ab}=\pi^{\langle ab\rangle} \delta q_{ab}+\frac{c_1}{2}\sqrt{q}q^{ab}\delta q_{ab}=\pi^{\langle ab\rangle} \delta q_{ab}+c_1\delta(\sqrt{q})~.
\ee
Both the symplectic two-form and the physical Hamiltonian are unaffected by a non-zero value for $c_1$, and therefore the dynamics is equivalent to the $c_1=0$ case.

The canonical action for the singular case the  reads 
\be\label{singular_action}
S^{\rm GF}_{\rm sing}[q,\pi]  = \int \de t\, \de^3x \left[ \bar{\pi}^{  ab } \dot{\bar{q}}_{ab} - {\cal H}^{\rm P}_{\rm sing} -\bar{\pi}^{ ab } {\cal L}_{N} \bar{q}_{ ab} \right]~,
\ee
where $\bar{q}_{ab}$ is any solution of Eq.~\eqref{Eq:TracelessPI_TildeQ}.
 
We note that the expression for the physical Hamiltonian (\ref{Hsing}) still depends on $q_{ab}$ through each of its terms. Therefore the variation of  ${\cal H}^{\rm P}_{\rm sing}$ is of the form 
\be
\delta {\cal H}^{\rm P}_{\rm sing} = (\cdots)_{ab} \delta \bar{\pi}^{ab}  + (\cdots)^{ab} \delta \bar{q}_{ab} + (\cdots)\delta q.
\ee
The first two terms give the Hamilton equations for the canonical variables $(\bar{q}_{ab},\bar{\pi}^{ab} )$, whereas the last term  gives an additional equation: the coefficient of $\delta q$ must vanish.  Let us compute this term:
\be
\frac{\delta {\cal H}^{\rm P}_{\rm sing}}{\delta q} = \frac{\delta {\cal H}^{\rm P}_{\rm sing}}{\delta q_{ab}} \frac{\delta q_{ab}}{\delta q} 
= \frac{1}{q}\  q^{ab}  \frac{\delta {\cal H}^{\rm P}_{\rm sing}}{\delta q_{ab}} = \frac{1}{6q}\  {\cal H}^{\rm P}_{\rm sing}~.
\ee
 Thus we find that for this special case the physical Hamiltonian must vanish.  This is consistent with the analysis presented in the Appendix~\ref{Sec:Appendix}, where it is shown that $p_\phi$ must be zero.\footnote{Recall that, after gauge fixing, $p_\phi$ is no longer a phase-space variable but instead becomes a function of the remaining canonical variables, and coincides with the negative of the physical Hamiltonian.} This requirement constitutes a restriction on the initial conditions (in the cosmological case considered in Section~\ref{Sec: Cosmo_Examples} it amounts to a vanishing energy density for the dust component).
Lastly, for $f(\chi)=\frac{1}{3}\chi^2$,  i.e. $c_0=c_1=0$ in (\ref{specialf}), we see from (\ref{pi_sing}) that $\bar{\pi}=0$, and the last term in the physical Hamiltonian \eqref{Hsing} becomes zero. Thus, although this physical Hamiltonian  becomes the same as in GR coupled to a dust fluid \cite{Husain:2011tk}, the theory is in fact not the same due to the singular symplectic structure (\ref{GFaction_final2}), as discussed in the last paragraph: it has one less degree of freedom.  

\end{itemize}

To summarize this section, we obtained the action and physical Hamiltonian for mimetic gravity in the time gauge $\phi=t$ as a theory of only the metric $q_{ab}$ and its conjugate momentum $\bar{\pi}^{ab}$. We also commented in detail on the degenerate case of $f(\chi)$ (\ref{specialf}). We conclude this section with a few comments. 

The Dirac bracket~\eqref{Dirac_bracket} has been previously obtained in Ref.~\cite{Bodendorfer:2017bjt} following the Dirac algorithm and imposing $D_a\phi=0$ only after computing the inverse of the Dirac matrix. However in that work the status of the condition $D_a\phi=0$ within the canonical theory was not clear. Therefore the relation between  the Dirac bracket and the symplectic structure of the reduced phase-space could not be fully established. We have shown that if the more restrictive condition $\phi=t$ is imposed as a canonical gauge fixing condition the reduced phase-space can be easily obtained after solving the (second class) constraints \eqref{Constraints_AuxiliaryFields} algebraically. Similar considerations apply to any gauge-fixing condition of the type $\phi=\phi(t)$, as long as $\phi(t)$ is invertible;  we focused on $\phi=t$ since it is a natural choice that leads to a simple form for the physical Hamiltonian. The condition $D_a\phi=0$ could also be imposed as a canonical gauge-fixing, since it is second class with the Hamiltonian constraint, but it does not fully fix the gauge and leaves  the lapse as an arbitrary function of time.

\section{Linearized Theory around Minkowski}\label{Sec:Linearized}

 We now derive the linearized equations of motion about the flat spacetime using the action (\ref{GFaction_final2}); it is algebraically easier to perform the perturbative expansion using the variables ($q_{ab},\pi^{ab}$) and convert to the canonically conjugated variables  ($q_{ab},\bar{\pi}^{ab}$) at a later stage. We use the approach developed in Ref.~\cite{Ali:2015ftw}, where the  case $f(\chi)=0$ is studied.  
 
Assuming $f(0)=0$ (i.e.,~vanishing cosmological constant) the background solution is:
\be
q_{ab}^{(0)}= e_{ab}~,~~\pi^{ab(0)}=0~,~~N^{a (0)} = 0~.
\ee
  Without loss of generality, it is also convenient to set $f^\prime(0)=0$, which can always be achieved by means of a canonical transformation (in fact, a non-zero value for $f^\prime(0)$ only changes the symplectic potential~\eqref{symplectic_potential_reduced} by an exact differential). It follows  from Eq.~(\ref{chi_momentumtrace}) that $\chi =0$. We introduce the expansion of the fields
 \bea\label{Eq:PertExp}
 q_{ab}(t,\vec{x}) &=& e_{ab} + h_{ab}(t,\vec{x})~,\\
\pi^{ab}(t,\vec{x}) &=&   0+p^{ab}(t,\vec{x})~,\nonumber\\
N^{a}(t,\vec{x}) &= & 0 + \xi^{a}(t,\vec{x})~,\nonumber
 \eea
 Equation~\eqref{chi_momentumtrace} then gives, to first order
 \be\label{Eq:PertSolChi}
 \chi\simeq-p~,
 \ee
 where $p=e_{ab}p^{ab}$.
 %Equation~\eqref{chi_momentumtrace} then gives, including terms up to second order in the perturbations
% \be\label{Eq:PertSolChi}
 %\chi\simeq-(p+h_{ab}p^{ap}-\frac{1}{2}hp)~,
 %\ee
 %where $h$ and $p$ denote the trace of the metric and momentum perturbations, respectively.
 We substitute the expansions~\eqref{Eq:PertExp} and Eq.~\eqref{Eq:PertSolChi} into the action (\ref{GFaction_final2}) and expand to second order in the perturbations to obtain, up to surface terms
\be
S^{(2)}\equiv\int{\de t\, \de^{3}x\left[\left(p^{ab}-\frac{1}{2}f^{\prime\prime}(0)\,p\,e^{ab}\right)\dot{h}_{ab} - {\cal H}^{(2)}_{\rm P} - \xi^{a}\bar{C}_{a}^{(1)}\right]}~, \label{pert-action0}
\ee
where
\begin{align}
\mathcal{H}^{(2)}_P &= 2 \left( p^{ab}p_{ab} -\frac{1}{2} p^2 \right) 
- \frac{h}{4 }\left(\p_a\p_b h^{ab} - \frac{1}{2} \p^2h  \right) 
+ \frac{h^{ab}}{4 } \left(\p_b\p^c h_{ca} -\frac{1}{2} \p^2 h_{ab}  \right)+\frac{1}{2}f^{\prime\prime}(0)p^2\\ 
\bar{C}_{a}^{(1)} &= -2 \p^b \left( p_{ab} -\frac{1}{2}f^{\prime\prime}(0)\,p\,e_{ab} \right)~.
\end{align}

Now we introduce a new momentum variable
\be
\bar{p}^{ab}=p^{ab}-\frac{1}{2}f^{\prime\prime}(0)\,p\,e^{ab}~,
\ee
 which is canonically conjugate to the metric perturbation $h_{ab}$. Thus, we have, inverting this equation $p^{ab}=\bar{p}^{ab}+\gamma\bar{p}\,e^{ab}$, where we used the notation $\gamma=\tfrac{1}{2}f^{\prime\prime}(0)/(1-\tfrac{3}{2}f^{\prime\prime}(0))$.
After this transformation, the action reads
\be
S^{(2)}=\int{\de t\, \de^{3}x\left[\bar{p}^{ab}\dot{h}_{ab} - \bar{{\cal H}}^{(2)}_{\rm P} - \xi^{a}\bar{C}_{a}^{(1)}\right]}~, \label{pert-action}
\ee
with
\begin{align}
\mathcal{H}^{(2)}_P &= 2 \left( \bar{p}^{ab}\bar{p}_{ab} -\frac{1}{2} \bar{p}^2 \right) 
- \frac{h}{4 }\left(\p_a\p_b h^{ab} - \frac{1}{2} \p^2h  \right) 
+ \frac{h^{ab}}{4 } \left(\p_b\p^c h_{ca} -\frac{1}{2} \p^2 h_{ab}  \right)-\gamma\, \bar{p}^2\\ 
\bar{C}_{a}^{(1)} &= -2 \p^b \bar{p}_{ab} ~.
\end{align}

The second-order action is most easily analyzed in $3-$momentum space. For this purpose we introduce the Fourier expansions
\bea 
\label{eq:sfs}
 h_{ab} (t,\vec{x}) &=& \frac{1}{(2\pi)^{3/2}}\int{\de^{3}k\left[e^{i\vec{k}.\vec{x}}\,M_{ab}^I(\vec{k}) h_I(t,\vec{k})\right]}, \\
\bar{p}^{ab} (t,\vec{x}) &=&  \frac{1}{(2\pi)^{3/2}} \int{\de^{3}k\left[e^{i\vec{k}.\vec{x}}\,M^{ab}_I(\vec{k})p^I(t,\vec{k})\right]},\\
\xi^{a} (t,\vec{x}) &=&  \frac{1}{(2\pi)^{3/2}}\int{\de^{3}k\left[e^{i\vec{k}.\vec{x}}\,\tilde{\xi}^{a}(t,\vec{k})\right]}.
 \eea
 Here the matrices $M_{ab}^I(\vec{k}), \ I=1\cdots 6$ (to be defined below) form a time-independent basis for $3\times 3$ real symmetric matrices, that give a decomposition of the gravitational  phase-space variables into the canonical set $(h^I,p_I)$. The matrices $M^I$  must satisfy the orthogonality condition
 \be
   \text{Tr} (M^I(\vec{k})M^J(\vec{k})) = M_{ab}^I(\vec{k}) M^{ab J}(\vec{k}) = \delta^{I J}~. \label{orthogM}
 \ee
  The matrices $M^I(\vec{k})$ are defined using the unit vector $\kappa^a = k^a/|k|$ and the eigenvectors $\epsilon^a_\pm =(\eps^a_1 \pm i\eps^a_2)/\sqrt{2}$ of rotations about the $\kappa^a$ axis (where $\eps_1^a,\eps_2^a$ are coordinate unit vectors for the flat metric $e_{ab}$, and together with $\kappa^a$ they form a right-handed basis). These fall into the following cases that respectively transform as scalars, tensors, and vectors under the rotation group:
 \bea
 M_{1}^{ab}(\vec{k}) &=& \frac{1}{\sqrt{3}}\ e^{ab},   \quad  \quad  \quad  \quad  \quad  \quad  \ 
 M_{2}^{ab}(\vec{k}) = \sqrt{\frac{3}{2}} \left( \kappa^a\kappa^b - \frac{1}{3}  e^{ab} \right)~,\nn\\
M_{3}^{ab}(\vec{k}) &=& \frac{i}{\sqrt{2}}\left(\eps_{-}^{a}\eps_{-}^{b}  - \eps_{+}^{a}\eps_{+}^{b}\right), \quad 
M_{4}^{ab}(\vec{k}) = \frac{1}{\sqrt{2}}\left(\eps_{-}^{a}\eps_{-}^{b}  + \eps_{+}^{a}\eps_{+}^{b}\right)~,\nn\\
 M_{5}^{ab}(\vec{k}) &=& i\left(\eps_{-}^{(a}\kappa^{b)}  - \eps_{+}^{(a}\kappa^{b)}\right),\quad \ \   
 M_{6}^{ab}(\vec{k}) = \eps_{-}^{(a}\kappa^{b)}  + \eps_{+}^{(a}\kappa^{b)}~, 
\eea
and satisfy the properties
\bea
 e^{ab}M_{ab}^I(\vec{k}) &=&0, \ \ I=2\cdots 6~; \nn\\
 \kappa^aM_{ab}^I(\vec{k}) &=& 0,   \ \  I=3,4~; \nn \\
 \kappa^a \kappa^bM_{ab}^I(\vec{k}) &=& 0, \ \ I = 5,6~. 
 \eea
  We also note that the matrices $M_{3}^{ab}(\vec{k})$ and $M_{6}^{ab}(\vec{k})$ are odd under the parity transformation $\vec{k}\to-\vec{k}$, whereby $\epsilon^a_\pm\to\epsilon^a_\mp$; the remaining matrices are parity-even. We express this property as
 \be
 M_{I}^{ab}(-\vec{k})=(-1)^{1+P(I)} M_{I}^{ab}(\vec{k})~,
 \ee
 where $P(I)=-1$ only for $I=3,\,6$, and $P(I)=1$ otherwise.

Reality of the real-space fields in Eqs.~\eqref{eq:sfs} implies that the Fourier coefficients must satisfy the following conditions:
 \be
 \left(h^I(t,\vec{k})\right)^{*}=(-1)^{1+P(I)}h^I(t,-\vec{k})~,~~\left(p^I(t,\vec{k})\right)^{*}=(-1)^{1+P(I)}p^I(t,-\vec{k})~,~~\left(\tilde{\xi}^{a}(t,\vec{k})\right)^{*}=\tilde{\xi}^{a}(t,-\vec{k})~.
 \ee
 
 %The properties above ensure that the symplectic structure is preserved when the canonical action for perturbations \eq{pert-action} is written in $k-$space, i.e.
 The properties above imply that the symplectic term in the canonical action for perturbations \eq{pert-action} in $k-$space reads as
 \be
 \label{hpbasis}
 \int \de^3x\, \de t \ \bar{p}^{ab}\dot{h}_{ab}=\int \de^3k\, \de t\  p^I(t,\vec{k})^{*}\,\dot{h}_I (t,\vec{k})~,
 \ee
 whence we read off the fundamental Poisson brackets
\be
\{h_I(\vec{k}),p^J(\vec{k}^{\prime})^{*}\}=\{h_I(\vec{k})^{*},p^J(\vec{k}^{\prime})\}=\delta_I^J\delta(\vec{k}-\vec{k}^{\prime})~.
\ee
 
 The perturbation of the shift vector may also be decomposed into its transverse ($\eps_1,\eps_2$) and longitudinal ($\kappa^a$) components as   
 \be
 \tilde{\xi}^{a}(t,\vec{k}) = \xi_{1}(t,\vec{k})\eps_{1}^{a} + \xi_{2}(t,\vec{k})\eps_{2}^{a} + \xi_{||}(t,\vec{k})\kappa^{a}.
 \ee 

The momentum space action then reads as
\be
S^{(2)}=\int{\de t\, \de^{3}k\left[p^I(\vec{k})^{*}\,\dot{h}_I(\vec{k})  - \tilde{H}^{(2)}_P(\vec{k}) - i\tilde{\xi}^{a}(\vec{k})^{*}\tilde{C}_{a}(\vec{k})\right]}~.
\label{k-act}
\ee
(Here and in the following we only indicate the momentum dependence and omit the time dependence in order to make the notation lighter.)
The second order Hamiltonian splits into a sum of three contributions $\tilde{H}^{(2)}_P = H^S + H^V +H^T$ (corresponding, respectively,  to scalars, vectors and tensors), given by (after suitable symmetrization over momenta $\vec{k}$ and $-\vec{k}$)
\begin{subequations}
\begin{align}
\label{HS}
H^S(\vec{k})  &=  2  \left(|p_2(\vec{k})|^2 -\frac{1}{2} |p_1(\vec{k})|^2  \right) - \frac{k^2}{12} \left| h_1(\vec{k})-\frac{1}{\sqrt{2}}h_2(\vec{k})\right|^2-3\gamma\, |p_1(\vec{k})|^2\\
  \label{HV}\nn\\
 H^V(\vec{k}) &=  2\left(|p_5(\vec{k})|^2 + |p_6(\vec{k})|^2\right)  %+  2c_1  \left(p_5h_5+p_6h_6 \right) +\frac{5}{4} c_1^2  \left(h_5^2 + h_6^2 \right)
 ,\\ \nn\\
 \label{HT}
 H^T(\vec{k}) &=  2\left(|p_3(\vec{k})|^2 + |p_4(\vec{k})|^2\right) %+  2c_1 \left(p_3h_3+p_4h_4\right) +\frac{1}{8}\left( 10c_1^2   +k^2\right) \left(h_3^2 + h_4^2 \right).
+ \frac{1}{8}k^2 \left( |h_3(\vec{k})|^2+ |h_4(\vec{k})|^2  \right)
\end{align}
\end{subequations}
 and the diffeomorphism constraint is
 \be\label{diff-comp}
 \tilde{C}_a(\vec{k})  =% -2 i k^b M^{I}_{ab}h_I(\vec{k})=
-2 k\left[   (p_{1}(\vec{k})+\sqrt{2}\,p_{2}(\vec{k}))\,\frac{\kappa_a}{\sqrt{3}} +p_6(\vec{k})\,\eps_{1a}  +p_5(\vec{k})\, \eps_{2a}\right]~.
\ee
 
    \subsection{Partial gauge fixing: removal of vector modes}

At this stage it is useful to perform a gauge-fixing to remove the  vector modes. This involves imposing canonical gauge conditions on these modes and solving strongly the corresponding diffeomorphism constraint  components. The above decomposition reveals the convenient choice
\be
 h_5 = h_6 =0~. \label{vec-gauge}
\ee
These conditions are second class with the transverse component $C_\perp$ of the diffeomorphism constraint, 
\be
\{h_5, C_\perp\} =-2 k\, \eps_{2a}~,~~ \{h_6, C_\perp \} = -2 k\, \eps_{1a} ~, 
\ee
 unless $k=0$.  Since we are interested in propagating modes (where the diffeomorphism constraint  is not identically zero),  and in regions far from a potential singularity, these gauge choices are sufficient. The constraint $C_\perp=0$ is then solved by setting $p_5=p_6=0$. 

With this gauge-fixing the second-order Hamiltonian  $\tilde{H}^{(2)}$ and the linearized diffeomorphism constraint now reduce respectively  to 
\bea
\tilde{H}^{(2)} &=& H^S + H^T,\\
 C_\parallel &\equiv&  -\frac{2}{\sqrt{3}}k  (p_{1}+\sqrt{2}p_{2})  =0~. \label{sc-diff}
\eea
This remaining system gives the dynamics of the graviton and scalar modes, with residual gauge symmetry generated by $C_\parallel$. It is useful to note that the graviton sector phase-space variables $(h_3, p_3)$ and $(h_4,p_4)$ have vanishing  Poisson brackets with this constraint, and so are gauge-invariant to this order.

The perturbation of the three-dimensional curvature scalar is
\be
R^{(3)}= \left(\p_a\p_b h^{ab} - \p^2 h\right)~,
\ee 
and in Fourier space becomes   
\be\label{Eq:CurvaturePerturbation}
\tilde{R}^{(3)}=\frac{2k^2}{\sqrt{3}}   \left(h_{1}-\frac{h_{2}}{\sqrt{2}}\right)~.
\ee
We observe that this quantity Poisson-commutes with the diffeomorphism constraint
\be
\{ C_\parallel, \tilde{R}^{(3)}\} =0~.
\ee
Thus, the curvature perturbation is gauge-invariant under spatial diffeomorphisms.

The combination that appears in Eq.~\eqref{Eq:CurvaturePerturbation} is proportional to the Bardeen potential\footnote{The relation between the scalar perturbations $h_1$ and $h_2$ and more standard variables used in cosmological perturbation theory is: $h_1=-2\sqrt{3}\,\psi$,  $h_2=(2/\sqrt{3})\,k^2E$ (in Fourier space), using the conventions in Ref.~\cite{Riotto:2002yw}. The Bardeen potential is defined in terms of such variables as $\Psi_B=-\psi+\frac{1}{6}k^2E+\dot{a}(B-a\dot{E})$ (in our case we are expanding around Minkowski, and therefore $\dot{a}=0$ identically).}
\be
\Psi_{\rm B}=\frac{1}{2\sqrt 3}\left(h_{1}-\frac{h_{2}}{\sqrt{2}}\right)~,
\ee
and is therefore invariant under four-dimensional infinitesimal diffeomorphisms, as it is well-known. The curvature perturbation potential $\mathcal{R}$ in the $\delta\phi=0$ slicing is defined as (recall that we are working in the $\phi=t$ time gauge)
\be
\tilde{R}^{(3)}=4k^2\mathcal{R} ~,
\ee
whence it follows that $\mathcal{R}=\Psi_{\rm B}$.

%It is a well-known result in cosmological perturbation theory that $\mathcal{R}$ can be promoted to a fully gauge-invariant observable as $\mathcal{R}=\mathcal{R}|_{\delta\phi=0}+H\delta\phi/\dot{\phi}$. (Note that in the case of Minkowski $H=0$ and therefore $\mathcal{R}=\mathcal{R}|_{\delta\phi=0}$ in any gauge.)
%The gauge $\delta\phi=0$ has been used widely in the literature to study the dynamics of cosmological perturbations in this theory on a FLRW background, see e.g.~Refs.~\cite{}.
%It is worth remarking that in the theory at hand and for $f^{\prime\prime}(\chi)\neq 0$, the $\delta\phi=0$ gauge is not equivalent to the comoving gauge, which is instead defined by the condition that the velocity perturbation $\delta u=0$ be vanishing, see Ref.~\cite{}.

 \subsection{Tensor modes}
 The equations of motion for tensor modes are derived from the Hamiltonian~(\ref{HT}). Hamilton's equations read as
 \be
 \dot{h}_I(\vec{k}) =  \left\{h_I(\vec{k}), \int \de^3k\, H^T(\vec{k}^{\prime})\right\}, \quad \dot{p}_I(t,\vec{k}) = \left\{p_I(\vec{k}), \int \de^3k\, H^T(\vec{k}^{\prime})\right\}~,  \quad I=3,4
 \ee 
and lead to 
\be
\ddot{h}_I (\vec{k})+ k^2\, h_I(\vec{k}) =0, \quad I=3,4~.
\ee
This is consistent with the well-known result that the propagation of tensor perturbations in the theory at hand is the same as in GR. (We note that the Fourier components $h_I(\vec{k})$ are not all independent due to the reality conditions.)

%\mar{two consistency checks: 1) $c1\to0$ should give the usual GR graviton equation with $c^2=1$; 2) this should be compatible with the Minkowski limit of cosmological perturbation theory done previously by other authors (provided that the limit can be taken smoothly)}
  
\subsection{Scalar mode}
It is convenient to fix the residual gauge symmetry generated by the longitudinal component $C_\parallel$ of the diffeomorphism constraint. We choose the canonical gauge $h_2=0$, which is second-class with $C_\parallel$. We then solve the constraint $C_\parallel=0$, which implies $p_{1}+\sqrt{2}\,p_{2}=0$. Substituting these two conditions into Eq.~\eqref{HS}, the Hamiltonian for scalar perturbations becomes
\be\label{HS2}
H^S(\vec{k})  =   - \frac{k^2}{12} \left| h_1(\vec{k})\right|^2-3\gamma\, |p_1(\vec{k})|^2~.
\ee
In this gauge, the curvature perturbation is $\mathcal{R}=\frac{1}{2\sqrt 3} h_{1}$. Its conjugate momentum is therefore $\Pi_{\mathcal{R}}=2\sqrt 3\,p_1$. Substituting in Eq.~\eqref{HS2}, we finally obtain the Hamiltonian for the curvature perturbation
\be
H^S(\vec{k})  =  -\frac{\gamma}{4}\, |\Pi_{\mathcal{R}}(\vec{k})|^2  - k^2 \left| \mathcal{R}(\vec{k})\right|^2~.
\ee
This Hamiltonian is the flat-space limit of the result previously obtained in Ref.~\cite{Firouzjahi:2017txv}. As noted there, this Hamiltonian is never bounded from below; depending on the sign of $\gamma$, two distinct types of instabilities arise for $\gamma\neq0$: a ghost instability\footnote{At the quantum level, a ghost instability is responsible for vacuum decay, which was studied in Ref.~\cite{Ramazanov:2016xhp}. Ghosts can pose a problem at the classical level too, if the unstable modes are coupled to other fields (e.g., matter)~\cite{Sbisa:2014pzo}.}
(i.e.,~negative kinetic energy) for $\gamma>0$, and a gradient instability\footnote{We note that in the case of a gradient instability (i.e., imaginary sound speed) the equations governing the perturbations are elliptic PDEs rather than hyperbolic, which implies that the initial value problem is ill-posed and therefore uniqueness of the solution is lost in general~\cite{Ijjas:2018cdm}.}
 for $\gamma<0$ (see also Refs.~\cite{Ramazanov:2016xhp,Ijjas:2016pad}).
 In the  case $\gamma=0$, the dynamics of perturbations becomes ultra-local and the curvature perturbation is conserved.\footnote{Note that, if the theory is regarded as fundamental (as opposed to an effective classical theory), then at the quantum level small values of $\gamma$ are linked to a low strong coupling scale, as discussed in Ref.~\cite{Ramazanov:2016xhp}.}

The Hamilton equations read as
\be
\dot{\mathcal{R}}(\vec{k})=-\frac{\gamma}{2}\Pi_{\mathcal{R}}(\vec{k})~,~~\dot{\Pi}_{\mathcal{R}}(\vec{k})=2k^2\mathcal{R}(\vec{k})~,
\ee
which can be combined to give the wave equation
\be
\ddot{\mathcal{R}}(\vec{k})+\gamma k^2\,\mathcal{R}(\vec{k})=0 ~.
\ee
The quantity $\gamma$ is therefore interpreted as the sound speed. Note that scalar modes are superluminal for $|\gamma|>1$.

From the canonical action for scalar perturbations
\be
S^{(2)}=\int{\de t\, \de^{3}k\left[\Pi_{\mathcal{R}}(-\vec{k})\dot{\mathcal{R}}(\vec{k})  - H^S(\vec{k}) \right]}~,
\ee
we can write down the corresponding second-order action for gauge-invariant curvature perturbations in Lagrangian form
\be\label{Eq:CovariantSecondOrderActionScalar}
S^{(2)}=\int{\de t\, \de^{3}k\left[\frac{1}{\gamma}|\dot{\mathcal{R}}(\vec{k})|^2  +k^2 |\mathcal{R}(\vec{k})|^2\right]}~.
\ee

 As we noted earlier, the singular case $f^{\prime\prime}(\chi)=2/3$ is excluded from this perturbative analysis, since scalar perturbations are not part of phase space in this case. We observe that the sound speed $\gamma$ is divergent when the limit $f^{\prime\prime}(0)\to2/3$ is approached from either side (note that the sign of $\gamma$ depends on the direction of approach, which also determines the type of instability). This gives rise to a discontinuity in the number of degrees of freedom.

Another interesting limit is $f^{\prime\prime}(0)\to0$, whereby the sound speed $\gamma$ tends to zero and the dynamics of scalar perturbations becomes ultra-local. 
 In this limit, the number of degrees of freedom is preserved in the Hamiltonian theory, both at the perturbative level and non-perturbatively (see Section~\ref{Sec:DustTimeGauge}). This should be contrasted with the $\gamma\to0$ limit of the covariant action~\eqref{Eq:CovariantSecondOrderActionScalar}, which would naively appear to be singular.

\section{Symmetry reduced models}\label{Sec:SymmetryReduced}

We  apply symmetry reductions directly to the Hamiltonian formalism developed above by computing  the physical Hamiltonian  and dynamical equations  for cosmological and spherically symmetric spacetimes. While these cases have been studied in  the literature, this section serves merely as an illustration of our alternative method. In particular the equations for the spherically symmetric sector may be useful for studying generalizations of the Lema{\^i}tre-Tolman-Bondi (LTB) metrics, and for numerical studies of gravitational collapse with additional matter fields.   

\subsection{Cosmological spacetimes}\label{Sec: Cosmo_Examples}

For the $k=0$ FLRW model, the  ADM variables are parametrized by 
\be
 q_{ab}=a^2(t)\,e_{ab}~,~~\pi^{ab}=\frac{p_a(t)}{6a(t)}\,e^{ab}~.
\ee
This leads to the vanishing of the diffeomorphism constraint. Substituting this parametrization into the action~(\ref{GFaction_final2}) gives the symmetry-reduced gauge-fixed canonical action 
\be
S^{\rm GF} = V_0\int \de t\, \left[ \left(p_a +3a^2 f^{\prime}(\chi)\right) \dot{a} -{\cal H}^{\rm P}    \right] ~,
\ee
where $V_0$ is a fiducial comoving volume, and
\be\label{FLRW_physical_hamiltonian}
\mathcal{H}^{\rm P}=-\frac{p_a^2}{12a}+a^3 \tilde{\epsilon}(\chi)~,  
\ee
having defined
\be
\tilde{\epsilon}(\chi) = \left( \chi\, f^{\prime}(\chi) -f(\chi)\right), \quad \chi=-\frac{p_a}{2a^2}~.
\ee
The Dirac bracket (\ref{Dirac_bracket}) reduces to 
\be
\{a,p_a\}_{\star}=\left(1-\frac{3}{2}f^{\prime\prime}(\chi)\right)^{-1}.
\ee
This shows that for $f^{\prime\prime}= 2/3$ the Dirac bracket of $a$ and $p_a$ is ill-defined; this is due to the fact that in this special case $a$ and $p_a$ Poisson commute and therefore cease to be independent phase space variables, in agreement with our general discussion
%of the general case
in Section~\ref{Sec:DustTimeGauge}.  For $f^{\prime\prime}\ne 3/2$ Hamilton's equations are  
\begin{align}
\dot{a}&=\{a,\mathcal{H}^{\rm P}\}_{\star}=-\frac{p_a}{6a} ~,\label{FLRW_HamiltonEQ1}\\
\dot{p}_a&=\{p_a,\mathcal{H}^{\rm P}\}_{\star}=-\left(1-\frac{3}{2}f^{\prime\prime}(\chi)\right)^{-1}\left[\frac{1}{12}\frac{p_a^2}{a^2}\left(1-6f^{\prime\prime}(\chi)\right)+3a^2\tilde{\epsilon}(\chi)\right]~.\label{FLRW_HamiltonEQ2}
\end{align}
As already noted after the more general Eq.~\eqref{HamiltonEQS_general}, the relation between $\dot{a}$ and $p_a$ in \eqref{FLRW_HamiltonEQ1} is the standard ADM one, while \eqref{FLRW_HamiltonEQ2} includes deviations from the standard Friedmann dynamics obtained in GR. 

The  physical Hamiltonian (\ref{FLRW_physical_hamiltonian}) may be re-expressed using the energy density $\rho =  - \mathcal{H}^{\rm P}/a^3=p_\phi/a^3$ and the expansion scalar $\chi$ as 
\be
\frac{1}{3}\chi^2=\tilde{\epsilon}(\chi) + \rho~.
\ee
On shell, $\chi = 3H$  using Eq.~\eqref{FLRW_HamiltonEQ1}, and the last equation becomes  
\be\label{Eq:EffectiveFriedmann}
\left(\frac{\dot{a}}{a}\right)^2 = \frac{1}{3} \left[\rho +  \tilde{\epsilon}(\chi)\right]~.  
\ee
The first term on r.h.s is the energy density of pressureless dust, referred to in the literature as `mimetic dark matter'~\cite{Chamseddine:2013kea}. 
%(Recall that $\mathcal{H}^{\rm P}=-p_\phi$; therefore, $p_\phi$ gives the initial condition for the energy density of dust at $a=1$.)

We can alternatively obtain equivalent dynamics by working with the action (\ref{GFaction_final}), and the parametrization   
\be
q_{ab}=a^2(t)\,e_{ab}~,~~\bar{\pi}^{ab}=\frac{\bar{p}_a(t)}{6a(t)}\,e^{ab}~,
\ee
with the Poisson bracket  $\displaystyle \{ a,\bar{p}_a\}=1 $. The Hamiltonian becomes~(\ref{HPbar})
\be
  \bar{{\cal H}}^{\rm P}  = -\frac{\bar{p}_a^2}{12a}-a^3\left( f(\chi)-\frac{3}{4} \left(f^{\prime}(\chi)\right)^2 \right)
\ee
with $\chi$ given implicitly by Eq.~\eqref{pibar-chi}:
\be
\frac{\bar{p}_a}{2a^2} =  \frac{3}{2} f'(\chi) - \chi~.
\label{chibar}
\ee
These give the equation of motion
\be
\dot{a} = \{a, \bar{{\cal H}}^{\rm P} \}= -\frac{\bar{p}_a}{6a} -a^3 \{a,\chi\} f'(\chi) \left(1-\frac{3}{2}f''(\chi)\right)= -\frac{\bar{p}_a}{6a} +\frac{af'(\chi)}{2} =  -\frac{p_a}{6a}~,
\ee
where $\{a,\chi\}$ follows from  Eq.~(\ref{chibar}). The $\dot{\bar{p}}_a$ equation is similarly derived and can be transformed to Eq.~(\ref{FLRW_HamiltonEQ2}). Thus, either form of the action is suitable for deriving equations of motion, with a simple mapping between them given by the relation between $\pi^{ab}$ and $\bar{\pi}^{ab}$. A similar analysis is possible for homogeneous and anisotropic spacetimes, such as Bianchi models and Kantowski-Sachs. 

%%%%%%%%%%%%

\subsection{Spherically symmetric spacetime}

For this case we give a parametrization of the symmetry reduction starting from the action (\ref{GFaction_final}), since the computation is more streamlined in the variables ($q_{ab}, \bar{\pi}^{ab}$).  
\bea
q_{ab} &=& \Lambda(r,t)^2\ s_a s_b + \frac{R(r,t)^2}{r^2}\ ( e_{ab} - s_a
s_b)\\
\bar{\pi}^{ab} &=& \frac{\bar{P}_\Lambda(r,t)}{2\Lambda(r,t)}\ s^a s^b + \frac{r^2
\bar{P}_R(r,t)}{4R(r,t)}\
(e^{ab} - s^a s^b),
\label{reduc}
\eea
where $e_{ab}$ is the flat Euclidean three-metric and $s^a=\left(\frac{\pa}{\pa r}\right)^a$ is the radial vector having unit norm w.r.t.~$e_{ab}$.
%\mar{I find this last bit ambiguous since $s^a=\left(\frac{\pa}{\pa r}\right)^a$ is a unit vector w.r.t.~$e_{ab}$, but not w.r.t.~$q_{ab}$. I suggest the following change: ``where [...] and $s^a=\left(\frac{\pa}{\pa r}\right)^a$ is the radial vector having unit norm w.r.t.~$e_{ab}$.''}
The spatial line element is therefore
\be
\de\ell^2 = \Lambda^2(r,t) \de r^2 + R^2(r,t) \de\Omega^2.
\ee
With this form the symplectic term in (\ref{GFaction_final}) becomes
\be
\bar{\pi}^{ab}\dot{q}_{ab} %\longrightarrow
= \bar{P}_R\dot{R} + \bar{P}_\Lambda \dot{\Lambda}~,    
\ee 
and the  action (\ref{GFaction_final}) reduces to 
\be
S= 4\pi \int \de t\, \de r \left(\bar{P}_R\dot{R} + \bar{P}_\Lambda\dot{\Lambda} 
    - \bar{{\cal H}}^{\rm P}  - N^r \bar{C}_r\right)  +\ {\rm surface\ term},
\ee
where we have performed the angular integral. The surface term is necessary to  ensure that the action is functionally differentiable for specified fall-off conditions as $r \rightarrow \infty$.

Only in this section we use a prime to denote derivative w.r.t.~$r$. The physical Hamiltonian density and diffeomorphism constraints are %
\bea
 \label{Hsph}
\bar{{\cal H}}^{\rm P} &=& \frac{1}{R^2\Lambda}\left[\frac{1}{4}\ (\bar{P}_\Lambda \Lambda)^2 -
\frac{1}{2}(\bar{P}_\Lambda \Lambda)(\bar{P}_R R)\right]
\nn\\ 
&&+ \frac{1}{\Lambda^2}\left[ 2RR^{\prime\prime}\Lambda -2RR^{\prime}\Lambda^{\prime} +\Lambda (R^{\prime})^2 
 \right] -\Lambda R^2 F\left(\chi\right)~,\\\\
\bar{C}_r &=&   \bar{P}_R R^{\prime}  -\Lambda \bar{P}_\Lambda^{\prime}  = 0~,
\label{Csph}
\eea
where 
\be
F(\chi) = \left( f(\chi)-\frac{3}{4} \left(\frac{\de f}{\de \chi}\right)^2 \right)
\ee
and $\chi$ is given   by  
\be
\bar{\pi} = \bar{\pi}^{ab}q_{ab} = \frac{1}{2}\left( \Lambda \bar{P}_\Lambda + R \bar{P}_R \right) = \Lambda R^2\left(\frac{3}{2} \frac{\de f}{\de \chi} -\chi\right).
\label{chi-sph}
\ee
%\mar{NOTE: I have replaced $f'(\chi)$ with $\frac{\de f}{\de \chi}$ only in this section, since there is a risk of confusing the $\chi$-derivative with the $r$-derivative.}\\
At this stage we can fix the radial diffeomorphism freedom with the gauge $R(r,t)=r$. Solving the diffeomorphism constraint strongly for $\bar{P}_R$ , and substituting the result back into the action gives
\be
S_R^{\rm GF} = 4\pi \int \de t \, \de r \left[  \bar{P}_\Lambda\dot{\Lambda} - \bar{{\cal H}}^{\rm GF}_{\rm P} \right],
\ee
where 
\be
\bar{{\cal H}}^{\rm GF}_{\rm P}=  -\frac{\Lambda}{2}\left(\frac{\bar{P}_\Lambda^2}{2r}  \right)^{\prime}  + \Lambda   \left(\frac{r}{\Lambda^2} \right)^{\prime} 
    -\Lambda r^2  F\left(\chi\right)~.  
\ee
and (\ref{chi-sph}) becomes 
\be
  \frac{1}{2r^2} \left(r\bar{P}_\Lambda \right)^{\prime}  = \frac{3}{2} \frac{\de f}{\de \chi} -\chi.
\ee
The evolution  equations simplify to  
\bea
\dot{\Lambda} &=& \left\{ \Lambda, \int_0^\infty \de r\ \bar{{\cal H}}^{\rm GF}_{\rm P}\   \right\} = \frac{\bar{P}_\Lambda \Lambda^{\prime}}{2r} -\frac{r}{2}\left(\Lambda \frac{\de f}{\de \chi}\right)^{\prime}~~ ~,\label{Lambdadot}\\
\dot{\bar{P}}_\Lambda &=& \left\{ \bar{P}_\Lambda, \int_0^\infty \de r\  \bar{{\cal H}}^{\rm GF}_{\rm P}\   \right\} = \left(\frac{\bar{P}_\Lambda^2}{4r}  \right)^{\prime} - \frac{1}{\Lambda^2} +r^2  F\left(\chi\right)~,\label{PLambdadot}
\eea
%%%%%%%
\iffalse
\mar{I have replaced $\bar{r}$ with $r$ in the integral.}\\
\mar{Do we need to impose any regularity (or growth) conditions at $r=0$ to ensure that the integral of (5.25) exists?}\\
\mar{Shouldn't the last term in (5.27) be instead:
\be
- \int_0^\infty \de r\ \Lambda r^2 \frac{\de F}{\de \chi}\frac{\delta \chi}{\delta \bar{P}_\Lambda}=- \int_0^\infty \de r\ \Lambda r^2 \frac{\de f}{\de \chi}\left(1-\frac{3}{2}\frac{\de^2 f}{\de \chi^2}\right)\frac{\delta \chi}{\delta \bar{P}_\Lambda}~.
\ee
 $\frac{\delta \chi}{\delta \bar{P}_\Lambda}$ can be computed by taking the functional derivative of (5.26) w.r.t.~$ \bar{P}_\Lambda$, which gives:
 \be
 \frac{\delta \chi(r)}{\delta \bar{P}_\Lambda(r_1)}=\frac{1}{2r^2}\left(\frac{3}{2}\frac{\de^2 f}{\de \chi^2}-1\right)^{-1}\left(r\delta(r-r_1)\right)^{\prime}~.
 \ee
 After substituting and integrating by parts one finds
 \be
 -\frac{r}{2}\left(\Lambda \frac{\de f}{\de \chi}\right)^{\prime}~.
 \ee
 }\\
 \fi
 %%%%%%%
 where (\ref{chi-sph}) is used to simplify the r.h.s. of  (\ref{Lambdadot}). We note that these equations may be rewritten using the ADM momentum $\pi^{ab}$, where in the similar parametrization,    we have  $\bar{P}_\Lambda = P_\Lambda + \frac{\de f}{\de \chi}\,r^2$ using (\ref{new_momentum}). Equations~\eqref{Lambdadot} and \eqref{PLambdadot} represent the starting point for numerical investigations.
 
Among the features of  interest for  effective theories is the modification of the behaviour of apparent horizons.  These may be computed as  a function of phase space variables.  In the parametrization we are using, in the gauge $R=r$, the radially inward and outward null expansions $\theta_-$ and $\theta_+$ are given by (see Ref.~\cite{Husain:2004yy})
\be
\theta_\pm = \mp\ \frac{P_\Lambda}{2\Lambda} -\left(r^2\Lambda \right)^{\prime}~.   
\ee
Thus  for a solution $(\Lambda, P_\Lambda)$, $\theta_+=0$ gives the horizon equation 
\be
P_\Lambda = -2\Lambda \left(r^2\Lambda \right)^{\prime}.
\ee  
These equations may also be written in terms of $\bar{P}_\Lambda$. It is therefore clear that $f(\chi)$ affects horizon location and evolution. Certain choices of $f$ may not even permit horizon formation, in which case $\theta_+$ is never zero: this is a possibility that deserves further study. 

Matter fields can be easily included in this scheme. For instance, if a minimally coupled scalar field $\psi=\psi(r,t)$ with a potential $V(\psi)$ is included, its contributions to the physical Hamiltonian~\eqref{Hsph} and to the radial diffeomorphism constraint~\eqref{Csph} are, respectively
\be
\mathcal{H}_\psi=\frac{1}{2\Lambda r^2}\pi_{\psi}^2+\frac{r^2}{2\Lambda}(\psi^{\prime})^2+\Lambda r^2V(\psi)~,\quad
C_{r,\,\psi}=\pi_{\psi}\psi^{\prime}~.
\ee
Such effective models  would provide alternatives to several that have been studied in the literature  from various points of view, all of which introduce mechanisms for singularity avoidance; see e.g. \cite{Husain:2008tc,Ziprick:2010vb,Kreienbuehl:2010vc, Benitez:2020szx}. These works  in turn are attempts to extend  well-established results in classical gravitational collapse  in spherically symmetry \cite{Choptuik:1992jv}.

\section{Summary and Discussion}\label{Sec:Discussion}

Our main result is the derivation of the physical Hamiltonian of mimetic gravity in the gauge $\phi=t$; we showed that this provides a complete time gauge fixing free of Gribov ambiguities.  In all earlier work,  this condition was use to provide a partial solution of the equations of motion or as a convenient condition for simplify constraint algebra calculations; its implications for the canonical theory were not addressed.  The structure of the physical Hamiltonian we derive is interesting; its first term is identical in form to the Hamiltonian constraint of GR, and the second term is a function of the expansion scalar.  

The method we followed used the canonical action at the forefront. The gauge fixed action (\ref{GFaction29}) led directly to the identification of the modified symplectic structure (\ref{symplectic_twoform}) after  elimination of the auxiliary fields $\lambda$ and $\phi$. This provides a symplectic-geometric picture of the derivation of the Dirac brackets; in the conventional approach the latter  would follow from identifying the second class constraints and constructing the Dirac matrix. 

We paid particular attention to the special case $f''(\chi) = 2/3$, where we showed that the number of physical degrees of freedom reduces by one to give a theory of two metric degrees of freedom. In the appendix we showed that this reduction may be viewed as a consequence of a hidden gauge symmetry that arises only in the gauges $D_a\phi=0$; we also elaborate there on the case where this gauge is not fixed -- the resulting theory in any other gauge turns out to have three configuration degrees of freedom. This is a highly unusual circumstance which is likely the result of the special structure in these theories coming from the constraint $g^{ab}\p_a\phi\p_b\phi =-1$. 

As applications of our canonical analysis we developed a Hamiltonian perturbation theory about the Minkowski space solution, deriving the tensor and scalar mode equations. We showed that these reproduce, relatively simply, the results of covariant analyses, including the  degenerate case. The spherically symmetric equations we derived provide a useful testing ground for numerical studies of gravitational collapse, similar to that done for general relativity coupled to a scalar field \cite{Choptuik:1992jv}.  

As a final comment, the method we used would illuminate the canonical structure of other scalar-vector-tensor theories, especially if these contain a pressureless dust field, or equivalently any scalar field subject to a timelike gradient condition.

\appendix
\renewcommand{\theequation}{\thesection.\arabic{equation}}

\section{Analysis of the singular case $f''(\chi)= 2/3$}\label{Sec:Appendix}
 
We examine in detail the singular case $f(\chi)=\frac{1}{3}\chi^2$ without first imposing the time gauge fixing $\phi=t$.
We will see that the reduction in the number of degrees of freedom noted above arises due to the emergence of a new gauge symmetry for this special case.

Let us first note that the auxiliary field $\chi$ may be replaced in the Hamiltonian constraint (\ref{Hamiltonian constraint}) using its equation of motion $\chi = -3\beta/2$. This gives
\be\label{Eq:HamConstrSingular}
\mathcal{H}=\frac{2}{\sqrt{q}}\left(\pi_{ab}^2-\frac{1}{2}\pi^2\right)-\frac{\sqrt{q}}{2}R^{(3)}+\frac{p_\beta}{\sqrt{q}}\left(p_\phi-\frac{\lambda}{2}p_\beta\right)+\sqrt{q}\left[\frac{\lambda}{2}\left(1+q^{ab}D_a\phi D_b\phi\right)+q^{ab}D_a\beta D_b \phi+\frac{3}{4}\beta^2\right].
\ee
Secondly, the action (\ref{action}) gives $\beta=-f^{\prime}(\chi)=-\frac{2}{3}\chi$ and $\chi=-\Box\phi$. These imply  
\be
\Box\phi-\frac{3}{2}\beta=0
\ee
Varying the action with respect to $\lambda$ gives the ``mimetic constraint" 
\be\label{Eq:SolutionMimeticConstraint}
({\cal L}_n\phi)^2=1+q^{ab}  D_a \phi D_b\phi~.
\ee
Taken together, the last two equations may be written as a function of phase space variables as follows.  First, the Laplacian of the scalar field can be written in the ADM decomposition as 
\bea
\Box \phi &\coloneqq& g^{ab}\nabla_a\nabla_b \phi=(q^{ab}-n^a n^b)\nabla_a \left[( q_b^{\; c}-n_b n^c)\nabla_c \phi\right]\nn\\
&=&\triangle \phi-{\cal L}_n^2\phi-K{\cal L}_n\phi+N^{-1} q^{ab}  D_a N D_b\phi~.
\eea
The term ${\cal L}_n^2\phi$ in the last expression may be expanded using Eq.~\eqref{Eq:SolutionMimeticConstraint} by taking the positive and differentiating to obtain
\be
\begin{split}
{\cal L}_n^2 \phi=&\frac{({\cal L}_n q^{ab}) D_a \phi D_b\phi +2 q^{ab} D_a \phi {\cal L}_n (D_b\phi)}{2\sqrt{1+q^{ab}D_a\phi D_b\phi}}=\frac{({\cal L}_n q^{ab}) D_a \phi D_b\phi +2 q^{ab} D_a \phi D_b({\cal L}_n\phi)}{2\sqrt{1+q^{ab}D_a\phi D_b\phi}}\\
=& -\frac{K^{ab} D_a \phi D_b\phi}{\sqrt{1+q^{ab}D_a\phi D_b\phi}}+q^{ab} D_a \phi D_b \log\sqrt{1+q^{ab}D_a\phi D_b\phi}~.
\end{split}
\ee
Using the above and expressing the result in terms of the ADM momenta gives 
\be
\Box\phi=\triangle \phi+\frac{2\pi^{ab} D_a \phi D_b\phi+\pi}{\sqrt{q}\sqrt{1+q^{ab}D_a\phi D_b\phi}}-q^{ab} D_a \phi D_b \log\sqrt{1+q^{ab}D_a\phi D_b\phi}+N^{-1} q^{ab}  D_a N D_b\phi~.
\ee
Multiplying this equation by $N\sqrt{q}$, and using the fact that $p_\beta=\sqrt{q}\sqrt{1+q^{ab}D_a\phi D_b\phi}$ from the canonical action ~\eqref{canonical_action1}, and discarding a surface term  leads to the smeared functional
\be
\int \de^3 x\; N\sqrt{q}\,\Box \phi=\int \de^3 x\; N\sqrt{q}\, p_{\beta}^{-1} \left( 2\pi^{ab} D_a \phi D_b\phi+\pi- \sqrt{q}\,q^{ab} D_a \phi D_b \left(\frac{p_\beta}{\sqrt{q}}\right) \right)~.
\ee
Now, rescaling the lapse gives a smeared version of the equation  $\Box\phi-\frac{3}{2}\beta=0$:
\be
L[N]\coloneqq\int \de^3 x\; 2N \left( 2\pi^{ab} D_a \phi D_b\phi+\pi- \sqrt{q}\,q^{ab} D_a \phi D_b \left(\frac{p_\beta}{\sqrt{q}}\right) -\frac{3}{2}\beta p_{\beta}\right)=0~.
\ee
This is a constraint equation (the overall factor of 2 is introduced   for convenience). In the $\phi=t$ gauge this constraint gives $\pi=3 \beta p_{\beta}/2$.  $L[N]$ generates the following transformations:
\begin{align}\label{Eq:ActionHstar}
\{q_{ab},L[N]\}&=2N(q_{ab}+2D_a\phi D_b\phi)~,\\
\{\pi^{ab},L[N]\}&=-2N\left(\pi^{ab}- \sqrt{q}\,D^a \phi D^b \left(\frac{p_\beta}{\sqrt{q}}\right)\right)~,\\
\{\beta,L[N]\}&=-3N \beta+\frac{2}{\sqrt{q}} \pa_a\left(N\sqrt{q}\, q^{ab} \pa_b \phi \right)~,\\
\{p_\beta,L[N]\}&=3N p_\beta~.
\end{align}
 In  time gauges where $D_a\phi=0$, and only in these gauges, $L[N]$  generates conformal transformations (Weyl rescalings);\footnote{It is interesting to compare the constraint $L[N]$ and its action on the dynamical fields with the Weyl constraint in Brans-Dicke theory with conformal coupling studied in Ref.~\cite{Gielen:2018pvk}.} the metric $q_{ab}$ and the canonical momentum $\pi^{ab}$ have conformal weights $+2$ and $-2$, while $\beta$ and its conjugate momentum $p_\beta$ have the non-standard conformal weights $-3$ and $3$, respectively (this is due to the fact that $p_\beta$ is actually a scalar density). 
However, in a generic frame, $L[N]$ generates a ``disformal" transformation, as it is clear from Eq.~\eqref{Eq:ActionHstar}.

Let us now define
\be\label{Eq:PhiConstraint}
M[\omega]\coloneqq\int \de^3 x\,\omega \left(p_{\beta}-\sqrt{q}\sqrt{1+q^{ab}D_a\phi D_b\phi}\right) =0.
\ee
This is the smeared version of the constraint obtained by varying the canonical action w.r.t.~the Lagrange multiplier $\lambda$. Since we have eliminated the auxiliary field $\chi$ from the Hamiltonian constraint~\eqref{Eq:HamConstrSingular}, $L[\varepsilon]$ and $M[\omega]$ and  are the only constraints other than the Hamiltonian and diffeomorphism constraints on the space space $(q_{ab},\pi^{ab};\beta,p_\beta)$.\footnote{For comparison, see also Ref.~\cite{Bodendorfer:2017bjt}, where additional constraints arise due to the presence of the other auxiliary fields, which we have eliminated at an early stage.}. 

We note the Poisson brackets of $L,M$ with the diffeomorphism and Hamiltonian constraints. The following are immediate:
\be
\{M[\omega],C[{\vec N}]\}={\cal L}_{N}M[\omega]\approx0~,\quad
\{L[\varepsilon],C[{\vec N}]\}={\cal L}_{N}L[\varepsilon]\approx0,\quad
 \{\Phi[\omega],{\cal H}[N]\}\approx0.
 \ee
For the bracket $\{L[\varepsilon],{\cal H}[N]\}$   we must differentiate between two cases. If $D_a\phi\neq0$, we have
\be\label{Eq:HstarStability1}
\{L[\varepsilon],{\cal H}[N]\}\approx \int\de^3 x\, \varepsilon N  \left[\frac{3\lambda}{\sqrt{q}}\left(p_\beta^2-q-\frac{4}{3}q q^{ab}D_a\phi D_b\phi\right) + \dots \right]~.
\ee
To avoid generating new constraints we can fix $\lambda$ such that the r.h.s vanishes; the precise form of the remaining terms on the r.h.s. of Eq.~\eqref{Eq:HstarStability1}  is irrelevant for argument). However, for $D_a\phi=0$  we must keep all terms in the bracket; the result is
\be\label{Eq:HstarStability2}
\{L[\varepsilon],{\cal H}[N]\}\approx -3\int\de^3 x\, \varepsilon N p_\phi~,
\ee
 Therefore to generate no new constraints (for  the case  $D_a\phi=0$) we must impose $p_\phi\approx0$. Recalling that $-p_\phi$ is the physical Hamiltonian in the gauge  $\phi=t$ as shown in Section~\ref{Sec:DustTimeGauge}, this amounts to a restriction on the initial data. An analysis in the more general gauge $\phi=f(t)$, where $f(t)$ is an arbitrary function, proceeds in close analogy and leads to the same conclusion.

In order to better understand the difference between the cases $D_a\phi=0$ and $D_a\phi\ne0$, let us compute the algebra of these constraints:
\begin{align}
\{M[\omega_1],M[\omega_2]\}&=0~,\\
\{M[\omega],L[\varepsilon]\}&=3\int\de^3 x\, \omega \varepsilon \left(p_\beta-\frac{\sqrt{q}}{\sqrt{1+q^{ab}D_a\phi D_b\phi}}\right)\approx 3\int\de^3 x\, \omega \varepsilon \sqrt{q} \left(\alpha-\frac{1}{\alpha}\right)~,\\
\{ L[\varepsilon_1],L[\varepsilon_2]\}&=-3\int\de^3 x\, p_\beta q^{ab}D_a\phi \left(\varepsilon_1\pa_b \varepsilon_2- \varepsilon_2\pa_b \varepsilon_1\right)~.
\end{align}
where $\alpha=\sqrt{1+q^{ab}D_a\phi D_b\phi}$\;. This shows that the constraint algebra of $L$ and $M$ is not closed  unless  $D_a\phi=0$. If $D_a\phi=0$ this algebra reduces to
\be
\{M[\omega_1],M[\omega_2]\}=0~,~~
\{M[\omega],L[\varepsilon]\}=M\left[3\omega\epsilon\right]~,~~
\{L[\varepsilon_1],L[\varepsilon_2]\}=0~.
\ee
 Thus the algebra becomes first class if $D_a\phi=0$ is imposed. This is the ``hidden symmetry"  together with the initial data condition $p_\phi=0$ noted above.
 
 Let us summarize. Starting from the configuration space $(\phi,\beta,q_{ab})$ we find the following  
 \begin{itemize}
 \item $D_a\phi\ne0:$ the constraints $L,M$ are preserved under evolution provided  $\lambda$ is fixed; the algebra of $L$ and $M$ is second class so these constraints must be solved strongly.  Therefore the phase space has 2 less  degrees of freedom per space point, for a total of 6.
 \item $D_a\phi =0:$ the constraints $L,M$ are preserved under evolution provided $p_\phi=0$; the algebra of $L$ and $M$ is first class. Therefore there are 4 less phase space degrees of freedom per point, for a total of 4.
\end{itemize}
This shows that the singular case $f''(\chi) = 2/3$ actually yields two distinct theories: the gauge $D_a\phi=0$ reveals a new gauge symmetry, resulting in one less configuration degree of freedom.  
  
 %This implies that the counting of dynamical degrees of freedom must change depending on the gauge:\footnote{We recall that each first class constraint lowers the dimension of the reduced phase space by two, whereas second class constraints only reduce it by one unit.}
%we have three propagating degrees of freedom in foliations where $D_a\phi\neq0$, and just two in the class of foliations where $D_a\phi=0$. Hence different gauges describe different physics and general covariance is lost.
 
\bibliography{references}

\begin{thebibliography}{43}%
\makeatletter
\providecommand \@ifxundefined [1]{%
 \@ifx{#1\undefined}
}%
\providecommand \@ifnum [1]{%
 \ifnum #1\expandafter \@firstoftwo
 \else \expandafter \@secondoftwo
 \fi
}%
\providecommand \@ifx [1]{%
 \ifx #1\expandafter \@firstoftwo
 \else \expandafter \@secondoftwo
 \fi
}%
\providecommand \natexlab [1]{#1}%
\providecommand \enquote  [1]{``#1''}%
\providecommand \bibnamefont  [1]{#1}%
\providecommand \bibfnamefont [1]{#1}%
\providecommand \citenamefont [1]{#1}%
\providecommand \href@noop [0]{\@secondoftwo}%
\providecommand \href [0]{\begingroup \@sanitize@url \@href}%
\providecommand \@href[1]{\@@startlink{#1}\@@href}%
\providecommand \@@href[1]{\endgroup#1\@@endlink}%
\providecommand \@sanitize@url [0]{\catcode `\\12\catcode `\$12\catcode
  `\&12\catcode `\#12\catcode `\^12\catcode `\_12\catcode `\%12\relax}%
\providecommand \@@startlink[1]{}%
\providecommand \@@endlink[0]{}%
\providecommand \url  [0]{\begingroup\@sanitize@url \@url }%
\providecommand \@url [1]{\endgroup\@href {#1}{\urlprefix }}%
\providecommand \urlprefix  [0]{URL }%
\providecommand \Eprint [0]{\href }%
\providecommand \doibase [0]{http://dx.doi.org/}%
\providecommand \selectlanguage [0]{\@gobble}%
\providecommand \bibinfo  [0]{\@secondoftwo}%
\providecommand \bibfield  [0]{\@secondoftwo}%
\providecommand \translation [1]{[#1]}%
\providecommand \BibitemOpen [0]{}%
\providecommand \bibitemStop [0]{}%
\providecommand \bibitemNoStop [0]{.\EOS\space}%
\providecommand \EOS [0]{\spacefactor3000\relax}%
\providecommand \BibitemShut  [1]{\csname bibitem#1\endcsname}%
\let\auto@bib@innerbib\@empty
%</preamble>
\bibitem [{\citenamefont {Ashtekar}\ and\ \citenamefont
  {Singh}(2011)}]{Ashtekar:2011ni}%
  \BibitemOpen
  \bibfield  {author} {\bibinfo {author} {\bibfnamefont {Abhay}\ \bibnamefont
  {Ashtekar}}\ and\ \bibinfo {author} {\bibfnamefont {Parampreet}\ \bibnamefont
  {Singh}},\ }\bibfield  {title} {\enquote {\bibinfo {title} {{Loop Quantum
  Cosmology: A Status Report}},}\ }\href {\doibase
  10.1088/0264-9381/28/21/213001} {\bibfield  {journal} {\bibinfo  {journal}
  {Class. Quant. Grav.}\ }\textbf {\bibinfo {volume} {28}},\ \bibinfo {pages}
  {213001} (\bibinfo {year} {2011})},\ \Eprint {http://arxiv.org/abs/1108.0893}
  {arXiv:1108.0893 [gr-qc]} \BibitemShut {NoStop}%
\bibitem [{\citenamefont {Horava}(2009)}]{Horava:2009uw}%
  \BibitemOpen
  \bibfield  {author} {\bibinfo {author} {\bibfnamefont {Petr}\ \bibnamefont
  {Horava}},\ }\bibfield  {title} {\enquote {\bibinfo {title} {{Quantum Gravity
  at a Lifshitz Point}},}\ }\href {\doibase 10.1103/PhysRevD.79.084008}
  {\bibfield  {journal} {\bibinfo  {journal} {Phys. Rev. D}\ }\textbf {\bibinfo
  {volume} {79}},\ \bibinfo {pages} {084008} (\bibinfo {year} {2009})},\
  \Eprint {http://arxiv.org/abs/0901.3775} {arXiv:0901.3775 [hep-th]}
  \BibitemShut {NoStop}%
\bibitem [{\citenamefont {Chamseddine}\ and\ \citenamefont
  {Mukhanov}(2013)}]{Chamseddine:2013kea}%
  \BibitemOpen
  \bibfield  {author} {\bibinfo {author} {\bibfnamefont {Ali~H.}\ \bibnamefont
  {Chamseddine}}\ and\ \bibinfo {author} {\bibfnamefont {Viatcheslav}\
  \bibnamefont {Mukhanov}},\ }\bibfield  {title} {\enquote {\bibinfo {title}
  {{Mimetic Dark Matter}},}\ }\href {\doibase 10.1007/JHEP11(2013)135}
  {\bibfield  {journal} {\bibinfo  {journal} {JHEP}\ }\textbf {\bibinfo
  {volume} {11}},\ \bibinfo {pages} {135} (\bibinfo {year} {2013})},\ \Eprint
  {http://arxiv.org/abs/1308.5410} {arXiv:1308.5410 [astro-ph.CO]} \BibitemShut
  {NoStop}%
%%CITATION = ARXIV:1308.5410;%%
\bibitem [{\citenamefont {Chamseddine}\ and\ \citenamefont
  {Mukhanov}(2017)}]{Chamseddine:2016uef}%
  \BibitemOpen
  \bibfield  {author} {\bibinfo {author} {\bibfnamefont {Ali~H.}\ \bibnamefont
  {Chamseddine}}\ and\ \bibinfo {author} {\bibfnamefont {Viatcheslav}\
  \bibnamefont {Mukhanov}},\ }\bibfield  {title} {\enquote {\bibinfo {title}
  {{Resolving Cosmological Singularities}},}\ }\href {\doibase
  10.1088/1475-7516/2017/03/009} {\bibfield  {journal} {\bibinfo  {journal}
  {JCAP}\ }\textbf {\bibinfo {volume} {1703}},\ \bibinfo {pages} {009}
  (\bibinfo {year} {2017})},\ \Eprint {http://arxiv.org/abs/1612.05860}
  {arXiv:1612.05860 [gr-qc]} \BibitemShut {NoStop}%
%%CITATION = ARXIV:1612.05860;%%
\bibitem [{\citenamefont {Hojman}\ \emph {et~al.}(1976)\citenamefont {Hojman},
  \citenamefont {Kuchar},\ and\ \citenamefont {Teitelboim}}]{Hojman:1976vp}%
  \BibitemOpen
  \bibfield  {author} {\bibinfo {author} {\bibfnamefont {S.A.}\ \bibnamefont
  {Hojman}}, \bibinfo {author} {\bibfnamefont {K.}~\bibnamefont {Kuchar}}, \
  and\ \bibinfo {author} {\bibfnamefont {C.}~\bibnamefont {Teitelboim}},\
  }\bibfield  {title} {\enquote {\bibinfo {title} {{Geometrodynamics
  Regained}},}\ }\href {\doibase 10.1016/0003-4916(76)90112-3} {\bibfield
  {journal} {\bibinfo  {journal} {Annals Phys.}\ }\textbf {\bibinfo {volume}
  {96}},\ \bibinfo {pages} {88--135} (\bibinfo {year} {1976})}\BibitemShut
  {NoStop}%
\bibitem [{\citenamefont {Brown}\ and\ \citenamefont
  {Kuchar}(1995)}]{Brown:1994py}%
  \BibitemOpen
  \bibfield  {author} {\bibinfo {author} {\bibfnamefont {J.David}\ \bibnamefont
  {Brown}}\ and\ \bibinfo {author} {\bibfnamefont {Karel~V.}\ \bibnamefont
  {Kuchar}},\ }\bibfield  {title} {\enquote {\bibinfo {title} {{Dust as a
  standard of space and time in canonical quantum gravity}},}\ }\href {\doibase
  10.1103/PhysRevD.51.5600} {\bibfield  {journal} {\bibinfo  {journal} {Phys.
  Rev. D}\ }\textbf {\bibinfo {volume} {51}},\ \bibinfo {pages} {5600--5629}
  (\bibinfo {year} {1995})},\ \Eprint {http://arxiv.org/abs/gr-qc/9409001}
  {arXiv:gr-qc/9409001} \BibitemShut {NoStop}%
\bibitem [{\citenamefont {Bojowald}\ and\ \citenamefont
  {Paily}(2012)}]{Bojowald:2011aa}%
  \BibitemOpen
  \bibfield  {author} {\bibinfo {author} {\bibfnamefont {Martin}\ \bibnamefont
  {Bojowald}}\ and\ \bibinfo {author} {\bibfnamefont {George~M.}\ \bibnamefont
  {Paily}},\ }\bibfield  {title} {\enquote {\bibinfo {title} {{Deformed General
  Relativity and Effective Actions from Loop Quantum Gravity}},}\ }\href
  {\doibase 10.1103/PhysRevD.86.104018} {\bibfield  {journal} {\bibinfo
  {journal} {Phys. Rev. D}\ }\textbf {\bibinfo {volume} {86}},\ \bibinfo
  {pages} {104018} (\bibinfo {year} {2012})},\ \Eprint
  {http://arxiv.org/abs/1112.1899} {arXiv:1112.1899 [gr-qc]} \BibitemShut
  {NoStop}%
\bibitem [{\citenamefont {Husain}\ and\ \citenamefont
  {Pawlowski}(2012)}]{Husain:2011tk}%
  \BibitemOpen
  \bibfield  {author} {\bibinfo {author} {\bibfnamefont {Viqar}\ \bibnamefont
  {Husain}}\ and\ \bibinfo {author} {\bibfnamefont {Tomasz}\ \bibnamefont
  {Pawlowski}},\ }\bibfield  {title} {\enquote {\bibinfo {title} {{Time and a
  physical Hamiltonian for quantum gravity}},}\ }\href {\doibase
  10.1103/PhysRevLett.108.141301} {\bibfield  {journal} {\bibinfo  {journal}
  {Phys. Rev. Lett.}\ }\textbf {\bibinfo {volume} {108}},\ \bibinfo {pages}
  {141301} (\bibinfo {year} {2012})},\ \Eprint {http://arxiv.org/abs/1108.1145}
  {arXiv:1108.1145 [gr-qc]} \BibitemShut {NoStop}%
\bibitem [{\citenamefont {Giesel}\ and\ \citenamefont
  {Thiemann}(2015)}]{Giesel:2012rb}%
  \BibitemOpen
  \bibfield  {author} {\bibinfo {author} {\bibfnamefont {Kristina}\
  \bibnamefont {Giesel}}\ and\ \bibinfo {author} {\bibfnamefont {Thomas}\
  \bibnamefont {Thiemann}},\ }\bibfield  {title} {\enquote {\bibinfo {title}
  {{Scalar Material Reference Systems and Loop Quantum Gravity}},}\ }\href
  {\doibase 10.1088/0264-9381/32/13/135015} {\bibfield  {journal} {\bibinfo
  {journal} {Class. Quant. Grav.}\ }\textbf {\bibinfo {volume} {32}},\ \bibinfo
  {pages} {135015} (\bibinfo {year} {2015})},\ \Eprint
  {http://arxiv.org/abs/1206.3807} {arXiv:1206.3807 [gr-qc]} \BibitemShut
  {NoStop}%
\bibitem [{\citenamefont {Rovelli}\ and\ \citenamefont
  {Smolin}(1994)}]{Rovelli:1993bm}%
  \BibitemOpen
  \bibfield  {author} {\bibinfo {author} {\bibfnamefont {Carlo}\ \bibnamefont
  {Rovelli}}\ and\ \bibinfo {author} {\bibfnamefont {Lee}\ \bibnamefont
  {Smolin}},\ }\bibfield  {title} {\enquote {\bibinfo {title} {{The Physical
  Hamiltonian in nonperturbative quantum gravity}},}\ }\href {\doibase
  10.1103/PhysRevLett.72.446} {\bibfield  {journal} {\bibinfo  {journal} {Phys.
  Rev. Lett.}\ }\textbf {\bibinfo {volume} {72}},\ \bibinfo {pages} {446--449}
  (\bibinfo {year} {1994})},\ \Eprint {http://arxiv.org/abs/gr-qc/9308002}
  {arXiv:gr-qc/9308002} \BibitemShut {NoStop}%
\bibitem [{\citenamefont {Alesci}\ \emph {et~al.}(2015)\citenamefont {Alesci},
  \citenamefont {Assanioussi}, \citenamefont {Lewandowski},\ and\ \citenamefont
  {M{\"a}kinen}}]{Alesci:2015wla}%
  \BibitemOpen
  \bibfield  {author} {\bibinfo {author} {\bibfnamefont {E.}~\bibnamefont
  {Alesci}}, \bibinfo {author} {\bibfnamefont {M.}~\bibnamefont {Assanioussi}},
  \bibinfo {author} {\bibfnamefont {J.}~\bibnamefont {Lewandowski}}, \ and\
  \bibinfo {author} {\bibfnamefont {I.}~\bibnamefont {M{\"a}kinen}},\
  }\bibfield  {title} {\enquote {\bibinfo {title} {{Hamiltonian operator for
  loop quantum gravity coupled to a scalar field}},}\ }\href {\doibase
  10.1103/PhysRevD.91.124067} {\bibfield  {journal} {\bibinfo  {journal} {Phys.
  Rev. D}\ }\textbf {\bibinfo {volume} {91}},\ \bibinfo {pages} {124067}
  (\bibinfo {year} {2015})},\ \Eprint {http://arxiv.org/abs/1504.02068}
  {arXiv:1504.02068 [gr-qc]} \BibitemShut {NoStop}%
\bibitem [{\citenamefont {Kluson}(2017)}]{Kluson:2017iem}%
  \BibitemOpen
  \bibfield  {author} {\bibinfo {author} {\bibfnamefont {J.}~\bibnamefont
  {Kluson}},\ }\bibfield  {title} {\enquote {\bibinfo {title} {{Canonical
  Analysis of Inhomogeneous Dark Energy Model and Theory of Limiting
  Curvature}},}\ }\href {\doibase 10.1007/JHEP03(2017)031} {\bibfield
  {journal} {\bibinfo  {journal} {JHEP}\ }\textbf {\bibinfo {volume} {03}},\
  \bibinfo {pages} {031} (\bibinfo {year} {2017})},\ \Eprint
  {http://arxiv.org/abs/1701.08523} {arXiv:1701.08523 [hep-th]} \BibitemShut
  {NoStop}%
\bibitem [{\citenamefont {Bodendorfer}\ \emph
  {et~al.}(2018{\natexlab{a}})\citenamefont {Bodendorfer}, \citenamefont
  {Sch{\"a}fer},\ and\ \citenamefont {Schliemann}}]{Bodendorfer:2017bjt}%
  \BibitemOpen
  \bibfield  {author} {\bibinfo {author} {\bibfnamefont {N.}~\bibnamefont
  {Bodendorfer}}, \bibinfo {author} {\bibfnamefont {Andreas}\ \bibnamefont
  {Sch{\"a}fer}}, \ and\ \bibinfo {author} {\bibfnamefont {J.}~\bibnamefont
  {Schliemann}},\ }\bibfield  {title} {\enquote {\bibinfo {title} {{Canonical
  structure of general relativity with a limiting curvature and its relation to
  loop quantum gravity}},}\ }\href {\doibase 10.1103/PhysRevD.97.084057}
  {\bibfield  {journal} {\bibinfo  {journal} {Phys. Rev.}\ }\textbf {\bibinfo
  {volume} {D97}},\ \bibinfo {pages} {084057} (\bibinfo {year}
  {2018}{\natexlab{a}})},\ \Eprint {http://arxiv.org/abs/1703.10670}
  {arXiv:1703.10670 [gr-qc]} \BibitemShut {NoStop}%
%%CITATION = ARXIV:1703.10670;%%
\bibitem [{\citenamefont {Golovnev}(2014)}]{Golovnev:2013jxa}%
  \BibitemOpen
  \bibfield  {author} {\bibinfo {author} {\bibfnamefont {Alexey}\ \bibnamefont
  {Golovnev}},\ }\bibfield  {title} {\enquote {\bibinfo {title} {{On the
  recently proposed Mimetic Dark Matter}},}\ }\href {\doibase
  10.1016/j.physletb.2013.11.026} {\bibfield  {journal} {\bibinfo  {journal}
  {Phys. Lett. B}\ }\textbf {\bibinfo {volume} {728}},\ \bibinfo {pages}
  {39--40} (\bibinfo {year} {2014})},\ \Eprint {http://arxiv.org/abs/1310.2790}
  {arXiv:1310.2790 [gr-qc]} \BibitemShut {NoStop}%
\bibitem [{\citenamefont {Chamseddine}\ \emph {et~al.}(2014)\citenamefont
  {Chamseddine}, \citenamefont {Mukhanov},\ and\ \citenamefont
  {Vikman}}]{Chamseddine:2014vna}%
  \BibitemOpen
  \bibfield  {author} {\bibinfo {author} {\bibfnamefont {Ali~H.}\ \bibnamefont
  {Chamseddine}}, \bibinfo {author} {\bibfnamefont {Viatcheslav}\ \bibnamefont
  {Mukhanov}}, \ and\ \bibinfo {author} {\bibfnamefont {Alexander}\
  \bibnamefont {Vikman}},\ }\bibfield  {title} {\enquote {\bibinfo {title}
  {{Cosmology with Mimetic Matter}},}\ }\href {\doibase
  10.1088/1475-7516/2014/06/017} {\bibfield  {journal} {\bibinfo  {journal}
  {JCAP}\ }\textbf {\bibinfo {volume} {06}},\ \bibinfo {pages} {017} (\bibinfo
  {year} {2014})},\ \Eprint {http://arxiv.org/abs/1403.3961} {arXiv:1403.3961
  [astro-ph.CO]} \BibitemShut {NoStop}%
\bibitem [{\citenamefont {Sebastiani}\ \emph {et~al.}(2017)\citenamefont
  {Sebastiani}, \citenamefont {Vagnozzi},\ and\ \citenamefont
  {Myrzakulov}}]{Sebastiani:2016ras}%
  \BibitemOpen
  \bibfield  {author} {\bibinfo {author} {\bibfnamefont {L.}~\bibnamefont
  {Sebastiani}}, \bibinfo {author} {\bibfnamefont {S.}~\bibnamefont
  {Vagnozzi}}, \ and\ \bibinfo {author} {\bibfnamefont {R.}~\bibnamefont
  {Myrzakulov}},\ }\bibfield  {title} {\enquote {\bibinfo {title} {{Mimetic
  gravity: a review of recent developments and applications to cosmology and
  astrophysics}},}\ }\href {\doibase 10.1155/2017/3156915} {\bibfield
  {journal} {\bibinfo  {journal} {Adv. High Energy Phys.}\ }\textbf {\bibinfo
  {volume} {2017}},\ \bibinfo {pages} {3156915} (\bibinfo {year} {2017})},\
  \Eprint {http://arxiv.org/abs/1612.08661} {arXiv:1612.08661 [gr-qc]}
  \BibitemShut {NoStop}%
\bibitem [{\citenamefont {Langlois}\ \emph {et~al.}(2017)\citenamefont
  {Langlois}, \citenamefont {Liu}, \citenamefont {Noui},\ and\ \citenamefont
  {Wilson-Ewing}}]{Liu:2017puc}%
  \BibitemOpen
  \bibfield  {author} {\bibinfo {author} {\bibfnamefont {David}\ \bibnamefont
  {Langlois}}, \bibinfo {author} {\bibfnamefont {Hongguang}\ \bibnamefont
  {Liu}}, \bibinfo {author} {\bibfnamefont {Karim}\ \bibnamefont {Noui}}, \
  and\ \bibinfo {author} {\bibfnamefont {Edward}\ \bibnamefont
  {Wilson-Ewing}},\ }\bibfield  {title} {\enquote {\bibinfo {title} {{Effective
  loop quantum cosmology as a higher-derivative scalar-tensor theory}},}\
  }\href {\doibase 10.1088/1361-6382/aa8f2f} {\bibfield  {journal} {\bibinfo
  {journal} {Class. Quant. Grav.}\ }\textbf {\bibinfo {volume} {34}},\ \bibinfo
  {pages} {225004} (\bibinfo {year} {2017})},\ \Eprint
  {http://arxiv.org/abs/1703.10812} {arXiv:1703.10812 [gr-qc]} \BibitemShut
  {NoStop}%
%%CITATION = ARXIV:1703.10812;%%
\bibitem [{\citenamefont {de~Cesare}(2019{\natexlab{a}})}]{deCesare:2019pqj}%
  \BibitemOpen
  \bibfield  {author} {\bibinfo {author} {\bibfnamefont {Marco}\ \bibnamefont
  {de~Cesare}},\ }\bibfield  {title} {\enquote {\bibinfo {title}
  {{Reconstruction of Mimetic Gravity in a Non-SingularBouncing Universe from
  Quantum Gravity}},}\ }\href {\doibase 10.3390/universe5050107} {\bibfield
  {journal} {\bibinfo  {journal} {Universe}\ }\textbf {\bibinfo {volume} {5}},\
  \bibinfo {pages} {107} (\bibinfo {year} {2019}{\natexlab{a}})},\ \Eprint
  {http://arxiv.org/abs/1904.02622} {arXiv:1904.02622 [gr-qc]} \BibitemShut
  {NoStop}%
%%CITATION = ARXIV:1904.02622;%%
\bibitem [{\citenamefont {de~Cesare}(2019{\natexlab{b}})}]{deCesare:2018cts}%
  \BibitemOpen
  \bibfield  {author} {\bibinfo {author} {\bibfnamefont {Marco}\ \bibnamefont
  {de~Cesare}},\ }\bibfield  {title} {\enquote {\bibinfo {title} {{Limiting
  curvature mimetic gravity for group field theory condensates}},}\ }\href
  {\doibase 10.1103/PhysRevD.99.063505} {\bibfield  {journal} {\bibinfo
  {journal} {Phys. Rev.}\ }\textbf {\bibinfo {volume} {D99}},\ \bibinfo {pages}
  {063505} (\bibinfo {year} {2019}{\natexlab{b}})},\ \Eprint
  {http://arxiv.org/abs/1812.06171} {arXiv:1812.06171 [gr-qc]} \BibitemShut
  {NoStop}%
%%CITATION = ARXIV:1812.06171;%%
\bibitem [{\citenamefont {de~Haro}\ \emph {et~al.}(2019)\citenamefont
  {de~Haro}, \citenamefont {Arest\'e~Sal\'o},\ and\ \citenamefont
  {Pan}}]{deHaro:2018sqw}%
  \BibitemOpen
  \bibfield  {author} {\bibinfo {author} {\bibfnamefont {Jaume}\ \bibnamefont
  {de~Haro}}, \bibinfo {author} {\bibfnamefont {Llibert}\ \bibnamefont
  {Arest\'e~Sal\'o}}, \ and\ \bibinfo {author} {\bibfnamefont {Supriya}\
  \bibnamefont {Pan}},\ }\bibfield  {title} {\enquote {\bibinfo {title}
  {{Limiting curvature mimetic gravity and its relation to Loop Quantum
  Cosmology}},}\ }\href {\doibase 10.1007/s10714-019-2534-1} {\bibfield
  {journal} {\bibinfo  {journal} {Gen. Rel. Grav.}\ }\textbf {\bibinfo {volume}
  {51}},\ \bibinfo {pages} {49} (\bibinfo {year} {2019})},\ \Eprint
  {http://arxiv.org/abs/1803.09653} {arXiv:1803.09653 [gr-qc]} \BibitemShut
  {NoStop}%
%%CITATION = ARXIV:1803.09653;%%
\bibitem [{\citenamefont {de~Cesare}\ and\ \citenamefont
  {Wilson-Ewing}(2019)}]{deCesare:2019suk}%
  \BibitemOpen
  \bibfield  {author} {\bibinfo {author} {\bibfnamefont {Marco}\ \bibnamefont
  {de~Cesare}}\ and\ \bibinfo {author} {\bibfnamefont {Edward}\ \bibnamefont
  {Wilson-Ewing}},\ }\bibfield  {title} {\enquote {\bibinfo {title} {{A
  generalized Kasner transition for bouncing Bianchi I models in modified
  gravity theories}},}\ }\href@noop {} {\  (\bibinfo {year} {2019})},\ \Eprint
  {http://arxiv.org/abs/1910.03616} {arXiv:1910.03616 [gr-qc]} \BibitemShut
  {NoStop}%
%%CITATION = ARXIV:1910.03616;%%
\bibitem [{\citenamefont {Wilson-Ewing}(2018)}]{Wilson-Ewing:2017vju}%
  \BibitemOpen
  \bibfield  {author} {\bibinfo {author} {\bibfnamefont {Edward}\ \bibnamefont
  {Wilson-Ewing}},\ }\bibfield  {title} {\enquote {\bibinfo {title} {{The loop
  quantum cosmology bounce as a Kasner transition}},}\ }\href {\doibase
  10.1088/1361-6382/aaab8b} {\bibfield  {journal} {\bibinfo  {journal} {Class.
  Quant. Grav.}\ }\textbf {\bibinfo {volume} {35}},\ \bibinfo {pages} {065005}
  (\bibinfo {year} {2018})},\ \Eprint {http://arxiv.org/abs/1711.10943}
  {arXiv:1711.10943 [gr-qc]} \BibitemShut {NoStop}%
%%CITATION = ARXIV:1711.10943;%%
\bibitem [{\citenamefont {Bodendorfer}\ \emph
  {et~al.}(2018{\natexlab{b}})\citenamefont {Bodendorfer}, \citenamefont
  {Mele},\ and\ \citenamefont {M{\"u}nch}}]{Bodendorfer:2018ptp}%
  \BibitemOpen
  \bibfield  {author} {\bibinfo {author} {\bibfnamefont {Norbert}\ \bibnamefont
  {Bodendorfer}}, \bibinfo {author} {\bibfnamefont {Fabio~M.}\ \bibnamefont
  {Mele}}, \ and\ \bibinfo {author} {\bibfnamefont {Johannes}\ \bibnamefont
  {M{\"u}nch}},\ }\bibfield  {title} {\enquote {\bibinfo {title} {{Is limiting
  curvature mimetic gravity an effective polymer quantum gravity?}}}\ }\href
  {\doibase 10.1088/1361-6382/aae74b} {\bibfield  {journal} {\bibinfo
  {journal} {Class. Quant. Grav.}\ }\textbf {\bibinfo {volume} {35}},\ \bibinfo
  {pages} {225001} (\bibinfo {year} {2018}{\natexlab{b}})},\ \Eprint
  {http://arxiv.org/abs/1806.02052} {arXiv:1806.02052 [gr-qc]} \BibitemShut
  {NoStop}%
%%CITATION = ARXIV:1806.02052;%%
\bibitem [{\citenamefont {de~Cesare}\ \emph {et~al.}(2020)\citenamefont
  {de~Cesare}, \citenamefont {Seahra},\ and\ \citenamefont
  {Wilson-Ewing}}]{deCesare:2020swb}%
  \BibitemOpen
  \bibfield  {author} {\bibinfo {author} {\bibfnamefont {Marco}\ \bibnamefont
  {de~Cesare}}, \bibinfo {author} {\bibfnamefont {Sanjeev~S.}\ \bibnamefont
  {Seahra}}, \ and\ \bibinfo {author} {\bibfnamefont {Edward}\ \bibnamefont
  {Wilson-Ewing}},\ }\bibfield  {title} {\enquote {\bibinfo {title} {{The
  singularity in mimetic Kantowski-Sachs cosmology}},}\ }\href {\doibase
  10.1088/1475-7516/2020/07/018} {\bibfield  {journal} {\bibinfo  {journal}
  {JCAP}\ }\textbf {\bibinfo {volume} {07}},\ \bibinfo {pages} {018} (\bibinfo
  {year} {2020})},\ \Eprint {http://arxiv.org/abs/2002.11658} {arXiv:2002.11658
  [gr-qc]} \BibitemShut {NoStop}%
\bibitem [{\citenamefont {Ben~Achour}\ \emph {et~al.}(2018)\citenamefont
  {Ben~Achour}, \citenamefont {Lamy}, \citenamefont {Liu},\ and\ \citenamefont
  {Noui}}]{BenAchour:2017ivq}%
  \BibitemOpen
  \bibfield  {author} {\bibinfo {author} {\bibfnamefont {Jibril}\ \bibnamefont
  {Ben~Achour}}, \bibinfo {author} {\bibfnamefont {Frederic}\ \bibnamefont
  {Lamy}}, \bibinfo {author} {\bibfnamefont {Hongguang}\ \bibnamefont {Liu}}, \
  and\ \bibinfo {author} {\bibfnamefont {Karim}\ \bibnamefont {Noui}},\
  }\bibfield  {title} {\enquote {\bibinfo {title} {{Non-singular black holes
  and the Limiting Curvature Mechanism: A Hamiltonian perspective}},}\ }\href
  {\doibase 10.1088/1475-7516/2018/05/072} {\bibfield  {journal} {\bibinfo
  {journal} {JCAP}\ }\textbf {\bibinfo {volume} {1805}},\ \bibinfo {pages}
  {072} (\bibinfo {year} {2018})},\ \Eprint {http://arxiv.org/abs/1712.03876}
  {arXiv:1712.03876 [gr-qc]} \BibitemShut {NoStop}%
%%CITATION = ARXIV:1712.03876;%%
\bibitem [{\citenamefont {Faddeev}(1969)}]{Faddeev_1969}%
  \BibitemOpen
  \bibfield  {author} {\bibinfo {author} {\bibfnamefont {L.~D.}\ \bibnamefont
  {Faddeev}},\ }\bibfield  {title} {\enquote {\bibinfo {title} {The feynman
  integral for singular lagrangians},}\ }\href {\doibase 10.1007/bf01028566}
  {\bibfield  {journal} {\bibinfo  {journal} {Theoretical and Mathematical
  Physics}\ }\textbf {\bibinfo {volume} {1}},\ \bibinfo {pages} {1--13}
  (\bibinfo {year} {1969})}\BibitemShut {NoStop}%
\bibitem [{\citenamefont {Barnich}\ \emph {et~al.}(1991)\citenamefont
  {Barnich}, \citenamefont {Henneaux},\ and\ \citenamefont
  {Schomblond}}]{Barnich:1991tc}%
  \BibitemOpen
  \bibfield  {author} {\bibinfo {author} {\bibfnamefont {Glenn}\ \bibnamefont
  {Barnich}}, \bibinfo {author} {\bibfnamefont {Marc}\ \bibnamefont
  {Henneaux}}, \ and\ \bibinfo {author} {\bibfnamefont {Christiane}\
  \bibnamefont {Schomblond}},\ }\bibfield  {title} {\enquote {\bibinfo {title}
  {{On the covariant description of the canonical formalism}},}\ }\href
  {\doibase 10.1103/PhysRevD.44.R939} {\bibfield  {journal} {\bibinfo
  {journal} {Phys. Rev. D}\ }\textbf {\bibinfo {volume} {44}},\ \bibinfo
  {pages} {939--941} (\bibinfo {year} {1991})}\BibitemShut {NoStop}%
\bibitem [{\citenamefont {Firouzjahi}\ \emph {et~al.}(2017)\citenamefont
  {Firouzjahi}, \citenamefont {Gorji},\ and\ \citenamefont
  {Hosseini~Mansoori}}]{Firouzjahi:2017txv}%
  \BibitemOpen
  \bibfield  {author} {\bibinfo {author} {\bibfnamefont {Hassan}\ \bibnamefont
  {Firouzjahi}}, \bibinfo {author} {\bibfnamefont {Mohammad~Ali}\ \bibnamefont
  {Gorji}}, \ and\ \bibinfo {author} {\bibfnamefont {Seyed~Ali}\ \bibnamefont
  {Hosseini~Mansoori}},\ }\bibfield  {title} {\enquote {\bibinfo {title}
  {{Instabilities in Mimetic Matter Perturbations}},}\ }\href {\doibase
  10.1088/1475-7516/2017/07/031} {\bibfield  {journal} {\bibinfo  {journal}
  {JCAP}\ }\textbf {\bibinfo {volume} {07}},\ \bibinfo {pages} {031} (\bibinfo
  {year} {2017})},\ \Eprint {http://arxiv.org/abs/1703.02923} {arXiv:1703.02923
  [hep-th]} \BibitemShut {NoStop}%
\bibitem [{\citenamefont {Langlois}\ \emph {et~al.}(2019)\citenamefont
  {Langlois}, \citenamefont {Mancarella}, \citenamefont {Noui},\ and\
  \citenamefont {Vernizzi}}]{Langlois:2018jdg}%
  \BibitemOpen
  \bibfield  {author} {\bibinfo {author} {\bibfnamefont {David}\ \bibnamefont
  {Langlois}}, \bibinfo {author} {\bibfnamefont {Michele}\ \bibnamefont
  {Mancarella}}, \bibinfo {author} {\bibfnamefont {Karim}\ \bibnamefont
  {Noui}}, \ and\ \bibinfo {author} {\bibfnamefont {Filippo}\ \bibnamefont
  {Vernizzi}},\ }\bibfield  {title} {\enquote {\bibinfo {title} {{Mimetic
  gravity as DHOST theories}},}\ }\href {\doibase
  10.1088/1475-7516/2019/02/036} {\bibfield  {journal} {\bibinfo  {journal}
  {JCAP}\ }\textbf {\bibinfo {volume} {02}},\ \bibinfo {pages} {036} (\bibinfo
  {year} {2019})},\ \Eprint {http://arxiv.org/abs/1802.03394} {arXiv:1802.03394
  [gr-qc]} \BibitemShut {NoStop}%
\bibitem [{\citenamefont {Ramazanov}\ \emph {et~al.}(2016)\citenamefont
  {Ramazanov}, \citenamefont {Arroja}, \citenamefont {Celoria}, \citenamefont
  {Matarrese},\ and\ \citenamefont {Pilo}}]{Ramazanov:2016xhp}%
  \BibitemOpen
  \bibfield  {author} {\bibinfo {author} {\bibfnamefont {S.}~\bibnamefont
  {Ramazanov}}, \bibinfo {author} {\bibfnamefont {F.}~\bibnamefont {Arroja}},
  \bibinfo {author} {\bibfnamefont {M.}~\bibnamefont {Celoria}}, \bibinfo
  {author} {\bibfnamefont {S.}~\bibnamefont {Matarrese}}, \ and\ \bibinfo
  {author} {\bibfnamefont {L.}~\bibnamefont {Pilo}},\ }\bibfield  {title}
  {\enquote {\bibinfo {title} {{Living with ghosts in Ho\v rava-Lifshitz
  gravity}},}\ }\href {\doibase 10.1007/JHEP06(2016)020} {\bibfield  {journal}
  {\bibinfo  {journal} {JHEP}\ }\textbf {\bibinfo {volume} {06}},\ \bibinfo
  {pages} {020} (\bibinfo {year} {2016})},\ \Eprint
  {http://arxiv.org/abs/1601.05405} {arXiv:1601.05405 [hep-th]} \BibitemShut
  {NoStop}%
\bibitem [{\citenamefont {Ijjas}\ \emph {et~al.}(2016)\citenamefont {Ijjas},
  \citenamefont {Ripley},\ and\ \citenamefont {Steinhardt}}]{Ijjas:2016pad}%
  \BibitemOpen
  \bibfield  {author} {\bibinfo {author} {\bibfnamefont {Anna}\ \bibnamefont
  {Ijjas}}, \bibinfo {author} {\bibfnamefont {Justin}\ \bibnamefont {Ripley}},
  \ and\ \bibinfo {author} {\bibfnamefont {Paul~J.}\ \bibnamefont
  {Steinhardt}},\ }\bibfield  {title} {\enquote {\bibinfo {title} {{NEC
  violation in mimetic cosmology revisited}},}\ }\href {\doibase
  10.1016/j.physletb.2016.06.052} {\bibfield  {journal} {\bibinfo  {journal}
  {Phys. Lett. B}\ }\textbf {\bibinfo {volume} {760}},\ \bibinfo {pages}
  {132--138} (\bibinfo {year} {2016})},\ \Eprint
  {http://arxiv.org/abs/1604.08586} {arXiv:1604.08586 [gr-qc]} \BibitemShut
  {NoStop}%
\bibitem [{\citenamefont {Ali}\ \emph {et~al.}(2016)\citenamefont {Ali},
  \citenamefont {Husain}, \citenamefont {Rahmati},\ and\ \citenamefont
  {Ziprick}}]{Ali:2015ftw}%
  \BibitemOpen
  \bibfield  {author} {\bibinfo {author} {\bibfnamefont {Masooma}\ \bibnamefont
  {Ali}}, \bibinfo {author} {\bibfnamefont {Viqar}\ \bibnamefont {Husain}},
  \bibinfo {author} {\bibfnamefont {Shohreh}\ \bibnamefont {Rahmati}}, \ and\
  \bibinfo {author} {\bibfnamefont {Jonathan}\ \bibnamefont {Ziprick}},\
  }\bibfield  {title} {\enquote {\bibinfo {title} {{Linearized gravity with
  matter time}},}\ }\href {\doibase 10.1088/0264-9381/33/10/105012} {\bibfield
  {journal} {\bibinfo  {journal} {Class. Quant. Grav.}\ }\textbf {\bibinfo
  {volume} {33}},\ \bibinfo {pages} {105012} (\bibinfo {year} {2016})},\
  \Eprint {http://arxiv.org/abs/1512.07854} {arXiv:1512.07854 [gr-qc]}
  \BibitemShut {NoStop}%
\bibitem [{\citenamefont {Husain}\ and\ \citenamefont
  {Saeed}(2020)}]{Husain:2020uac}%
  \BibitemOpen
  \bibfield  {author} {\bibinfo {author} {\bibfnamefont {Viqar}\ \bibnamefont
  {Husain}}\ and\ \bibinfo {author} {\bibfnamefont {Mustafa}\ \bibnamefont
  {Saeed}},\ }\bibfield  {title} {\enquote {\bibinfo {title} {{Cosmological
  perturbation theory with matter time}},}\ }\href@noop {} {\  (\bibinfo {year}
  {2020})},\ \Eprint {http://arxiv.org/abs/2007.06609} {arXiv:2007.06609
  [gr-qc]} \BibitemShut {NoStop}%
\bibitem [{\citenamefont {Riotto}(2003)}]{Riotto:2002yw}%
  \BibitemOpen
  \bibfield  {author} {\bibinfo {author} {\bibfnamefont {Antonio}\ \bibnamefont
  {Riotto}},\ }\bibfield  {title} {\enquote {\bibinfo {title} {{Inflation and
  the theory of cosmological perturbations}},}\ }\href@noop {} {\bibfield
  {journal} {\bibinfo  {journal} {ICTP Lect. Notes Ser.}\ }\textbf {\bibinfo
  {volume} {14}},\ \bibinfo {pages} {317--413} (\bibinfo {year} {2003})},\
  \Eprint {http://arxiv.org/abs/hep-ph/0210162} {arXiv:hep-ph/0210162}
  \BibitemShut {NoStop}%
\bibitem [{\citenamefont {Sbis{\`a}}(2015)}]{Sbisa:2014pzo}%
  \BibitemOpen
  \bibfield  {author} {\bibinfo {author} {\bibfnamefont {Fulvio}\ \bibnamefont
  {Sbis{\`a}}},\ }\bibfield  {title} {\enquote {\bibinfo {title} {{Classical
  and quantum ghosts}},}\ }\href {\doibase 10.1088/0143-0807/36/1/015009}
  {\bibfield  {journal} {\bibinfo  {journal} {Eur. J. Phys.}\ }\textbf
  {\bibinfo {volume} {36}},\ \bibinfo {pages} {015009} (\bibinfo {year}
  {2015})},\ \Eprint {http://arxiv.org/abs/1406.4550} {arXiv:1406.4550
  [hep-th]} \BibitemShut {NoStop}%
\bibitem [{\citenamefont {Ijjas}\ \emph {et~al.}(2019)\citenamefont {Ijjas},
  \citenamefont {Pretorius},\ and\ \citenamefont {Steinhardt}}]{Ijjas:2018cdm}%
  \BibitemOpen
  \bibfield  {author} {\bibinfo {author} {\bibfnamefont {Anna}\ \bibnamefont
  {Ijjas}}, \bibinfo {author} {\bibfnamefont {Frans}\ \bibnamefont
  {Pretorius}}, \ and\ \bibinfo {author} {\bibfnamefont {Paul~J.}\ \bibnamefont
  {Steinhardt}},\ }\bibfield  {title} {\enquote {\bibinfo {title} {{Stability
  and the Gauge Problem in Non-Perturbative Cosmology}},}\ }\href {\doibase
  10.1088/1475-7516/2019/01/015} {\bibfield  {journal} {\bibinfo  {journal}
  {JCAP}\ }\textbf {\bibinfo {volume} {01}},\ \bibinfo {pages} {015} (\bibinfo
  {year} {2019})},\ \Eprint {http://arxiv.org/abs/1809.07010} {arXiv:1809.07010
  [gr-qc]} \BibitemShut {NoStop}%
\bibitem [{\citenamefont {Husain}\ and\ \citenamefont
  {Winkler}(2005)}]{Husain:2004yy}%
  \BibitemOpen
  \bibfield  {author} {\bibinfo {author} {\bibfnamefont {Viqar}\ \bibnamefont
  {Husain}}\ and\ \bibinfo {author} {\bibfnamefont {Oliver}\ \bibnamefont
  {Winkler}},\ }\bibfield  {title} {\enquote {\bibinfo {title} {{Quantum black
  holes from null expansion operators}},}\ }\href {\doibase
  10.1088/0264-9381/22/21/L02} {\bibfield  {journal} {\bibinfo  {journal}
  {Class. Quant. Grav.}\ }\textbf {\bibinfo {volume} {22}},\ \bibinfo {pages}
  {L135--L142} (\bibinfo {year} {2005})},\ \Eprint
  {http://arxiv.org/abs/gr-qc/0412039} {arXiv:gr-qc/0412039} \BibitemShut
  {NoStop}%
\bibitem [{\citenamefont {Husain}(2008)}]{Husain:2008tc}%
  \BibitemOpen
  \bibfield  {author} {\bibinfo {author} {\bibfnamefont {Viqar}\ \bibnamefont
  {Husain}},\ }\bibfield  {title} {\enquote {\bibinfo {title} {{Critical
  behaviour in quantum gravitational collapse}},}\ }\href@noop {} {\  (\bibinfo
  {year} {2008})},\ \Eprint {http://arxiv.org/abs/0808.0949} {arXiv:0808.0949
  [gr-qc]} \BibitemShut {NoStop}%
\bibitem [{\citenamefont {Ziprick}\ and\ \citenamefont
  {Kunstatter}(2010)}]{Ziprick:2010vb}%
  \BibitemOpen
  \bibfield  {author} {\bibinfo {author} {\bibfnamefont {Jonathan}\
  \bibnamefont {Ziprick}}\ and\ \bibinfo {author} {\bibfnamefont {Gabor}\
  \bibnamefont {Kunstatter}},\ }\bibfield  {title} {\enquote {\bibinfo {title}
  {{Quantum Corrected Spherical Collapse: A Phenomenological Framework}},}\
  }\href {\doibase 10.1103/PhysRevD.82.044031} {\bibfield  {journal} {\bibinfo
  {journal} {Phys. Rev. D}\ }\textbf {\bibinfo {volume} {82}},\ \bibinfo
  {pages} {044031} (\bibinfo {year} {2010})},\ \Eprint
  {http://arxiv.org/abs/1004.0525} {arXiv:1004.0525 [gr-qc]} \BibitemShut
  {NoStop}%
\bibitem [{\citenamefont {Kreienbuehl}\ \emph {et~al.}(2012)\citenamefont
  {Kreienbuehl}, \citenamefont {Husain},\ and\ \citenamefont
  {Seahra}}]{Kreienbuehl:2010vc}%
  \BibitemOpen
  \bibfield  {author} {\bibinfo {author} {\bibfnamefont {Andreas}\ \bibnamefont
  {Kreienbuehl}}, \bibinfo {author} {\bibfnamefont {Viqar}\ \bibnamefont
  {Husain}}, \ and\ \bibinfo {author} {\bibfnamefont {Sanjeev~S.}\ \bibnamefont
  {Seahra}},\ }\bibfield  {title} {\enquote {\bibinfo {title} {{Modified
  general relativity as a model for quantum gravitational collapse}},}\ }\href
  {\doibase 10.1088/0264-9381/29/9/095008} {\bibfield  {journal} {\bibinfo
  {journal} {Class. Quant. Grav.}\ }\textbf {\bibinfo {volume} {29}},\ \bibinfo
  {pages} {095008} (\bibinfo {year} {2012})},\ \Eprint
  {http://arxiv.org/abs/1011.2381} {arXiv:1011.2381 [gr-qc]} \BibitemShut
  {NoStop}%
\bibitem [{\citenamefont {Benitez}\ \emph {et~al.}(2020)\citenamefont
  {Benitez}, \citenamefont {Gambini}, \citenamefont {Lehner}, \citenamefont
  {Liebling},\ and\ \citenamefont {Pullin}}]{Benitez:2020szx}%
  \BibitemOpen
  \bibfield  {author} {\bibinfo {author} {\bibfnamefont {Florencia}\
  \bibnamefont {Benitez}}, \bibinfo {author} {\bibfnamefont {Rodolfo}\
  \bibnamefont {Gambini}}, \bibinfo {author} {\bibfnamefont {Luis}\
  \bibnamefont {Lehner}}, \bibinfo {author} {\bibfnamefont {Steve}\
  \bibnamefont {Liebling}}, \ and\ \bibinfo {author} {\bibfnamefont {Jorge}\
  \bibnamefont {Pullin}},\ }\bibfield  {title} {\enquote {\bibinfo {title}
  {{Critical collapse of a scalar field in semiclassical loop quantum
  gravity}},}\ }\href {\doibase 10.1103/PhysRevLett.124.071301} {\bibfield
  {journal} {\bibinfo  {journal} {Phys. Rev. Lett.}\ }\textbf {\bibinfo
  {volume} {124}},\ \bibinfo {pages} {071301} (\bibinfo {year} {2020})},\
  \Eprint {http://arxiv.org/abs/2002.04044} {arXiv:2002.04044 [gr-qc]}
  \BibitemShut {NoStop}%
\bibitem [{\citenamefont {Choptuik}(1993)}]{Choptuik:1992jv}%
  \BibitemOpen
  \bibfield  {author} {\bibinfo {author} {\bibfnamefont {Matthew~W.}\
  \bibnamefont {Choptuik}},\ }\bibfield  {title} {\enquote {\bibinfo {title}
  {{Universality and scaling in gravitational collapse of a massless scalar
  field}},}\ }\href {\doibase 10.1103/PhysRevLett.70.9} {\bibfield  {journal}
  {\bibinfo  {journal} {Phys. Rev. Lett.}\ }\textbf {\bibinfo {volume} {70}},\
  \bibinfo {pages} {9--12} (\bibinfo {year} {1993})}\BibitemShut {NoStop}%
\bibitem [{\citenamefont {Gielen}\ \emph {et~al.}(2018)\citenamefont {Gielen},
  \citenamefont {de~Le{\'o}n~Ard{\'o}n},\ and\ \citenamefont
  {Percacci}}]{Gielen:2018pvk}%
  \BibitemOpen
  \bibfield  {author} {\bibinfo {author} {\bibfnamefont {Steffen}\ \bibnamefont
  {Gielen}}, \bibinfo {author} {\bibfnamefont {Rodrigo}\ \bibnamefont
  {de~Le{\'o}n~Ard{\'o}n}}, \ and\ \bibinfo {author} {\bibfnamefont {Roberto}\
  \bibnamefont {Percacci}},\ }\bibfield  {title} {\enquote {\bibinfo {title}
  {{Gravity with more or less gauging}},}\ }\href {\doibase
  10.1088/1361-6382/aadbd1} {\bibfield  {journal} {\bibinfo  {journal} {Class.
  Quant. Grav.}\ }\textbf {\bibinfo {volume} {35}},\ \bibinfo {pages} {195009}
  (\bibinfo {year} {2018})},\ \Eprint {http://arxiv.org/abs/1805.11626}
  {arXiv:1805.11626 [gr-qc]} \BibitemShut {NoStop}%
\end{thebibliography}%

\end{document}